\PassOptionsToPackage{table}{xcolor}
\documentclass[acmsmall, screen]{acmart}

\setcopyright{rightsretained}
\acmPrice{}
\acmDOI{10.1145/3622837}
\acmYear{2023}
\copyrightyear{2023}
\acmSubmissionID{oopslab23main-p307-p}
\acmJournal{PACMPL}
\acmVolume{7}
\acmNumber{OOPSLA2}
\acmArticle{261}
\acmMonth{10}
\received{2023-04-14}
\received[accepted]{2023-08-27}

\usepackage{epigraph}
\usepackage[nameinlink]{cleveref}
\usepackage{pifont} %
\usepackage{tikz}
\usepackage{subcaption} %
\usepackage[procnames]{listings}
\usepackage{threeparttable}
\usepackage{lipsum}
\usepackage{mathpartir}
\usepackage{xspace}
\usepackage{tabularx}
\usepackage{thmtools}
\usepackage{multirow}
\usepackage{multicol}
\usepackage{etoolbox}

\makeatletter
\renewenvironment{thmt@restatable}[3][]{%
  \thmt@toks{}%
  \stepcounter{thmt@dummyctr}%
  \long\def\thmrst@store##1{%
    \@xa\gdef\csname #3\endcsname{%
      \@ifstar{%
        \thmt@thisistheonefalse\csname thmt@stored@#3\endcsname
      }{%
        \thmt@thisistheonetrue\csname thmt@stored@#3\endcsname
      }%
    }%
    \@xa\long\@xa\gdef\csname thmt@stored@#3\@xa\endcsname\@xa{%
      \begingroup
      \ifthmt@thisistheone
      \else
        \@xa\protected@edef\csname the#2\endcsname{%
          \thmt@trivialref{thmt@@#3}{??}}%
        \ifcsname r@thmt@@#3\endcsname\else
          \G@refundefinedtrue
        \fi
        \@xa\let\csname c@#2\endcsname=\c@thmt@dummyctr
        \@xa\let\csname theH#2\endcsname=\theHthmt@dummyctr
        \let\label=\@gobble
        \let\ltx@label=\@gobble%
        \def\thmt@restorecounters{}%
        \@for\thmt@ctr:=\thmt@innercounters\do{%
          \protected@edef\thmt@restorecounters{%
            \thmt@restorecounters
            \protect\setcounter{\thmt@ctr}{\arabic{\thmt@ctr}}%
          }%
        }%
        \thmt@trivialref{thmt@@#3@data}{}%
      \fi
      \ifthmt@restatethis
        \thmt@restatethisfalse
      \else
        \csname #2\@xa\endcsname\ifx\@nx#1\@nx\else[{#1}]\fi
      \fi
      \ifthmt@thisistheone
         \thmt@rst@storecounters{#3}%
        \label{thmt@@#3}%
      \fi
      ##1%
      \csname end#2\endcsname
      \ifthmt@thisistheone\else\thmt@restorecounters\fi
      \endgroup
    }%
    \csname #3\@xa\endcsname\ifthmt@thisistheone\else*\fi
    \@xa\end\@xa{\@currenvir}
  }%
  \thmt@collect@body\thmrst@store
}{%
}
\makeatother

\newdimen\figrasterwd
\figrasterwd\textwidth

\lstdefinelanguage{pp}{
  morekeywords=[1]{let,in,if,else,then,rec,Some,None},
  morekeywords=[2]{<+>,<>,<|>,<$>,nest,nl,group,text,align,flatten,fill,flat,full,fail,cost,newline,reset,singleLine,break,hard_nl},
  morekeywords=[3]{},
  alsoletter={^|<>$+},
  morestring=[b]",
  morecomment=[l]{\#},
  morecomment=[s]{(*}{*)},
  moredelim=**[is][\color{white}]{(&}{&)},
}

\lstdefinelanguage{literal}{
  basicstyle=\small\ttfamily\color{green!50!black},
  numberstyle=\color{black},
  moredelim=**[is][\color{white}]{(&}{&)},
  moredelim=**[is][\color{black}]{(^}{^)},
}

\usetikzlibrary{decorations.pathreplacing,calc,shapes,positioning,tikzmark,snakes,external,intersections}

\usetikzlibrary{arrows,shadows,matrix,patterns,fit}

\usepackage{pgfplots}
\usepgfplotslibrary{colormaps} 

\newcommand{\cmark}{\ding{51}}
\newcommand{\xmark}{\ding{55}}

\newcommand{\Lexpressive}{\ensuremath{\Sigma_e}}
\newcommand{\Pexpressive}{\ensuremath{\Pi_e}}
\newcommand{\prettiester}{\textsc{PrettyExpressive}}

\newcommand{\tighten}{\looseness=-1}
\newcommand{\naive}{na\"ive}

\newcommand{\LayoutT}{\ensuremath{\mathcal{L}}}
\newcommand{\StrT}{\ensuremath{\mathsf{Str}}}

\DeclareMathOperator{\defn}{\Coloneqq}
\DeclareMathOperator{\vbar}{\, |\, }

\newcommand{\set}[1]{\{#1\}}

\newcommand{\pp}[1]{\lstinline[language=pp]{#1}}

\newcommand{\ghosted}[1]{{\color{gray} #1}}

\newcommand{\mT}{\ensuremath{\mathcal{M}}}

\newcommand{\mlw}{\ensuremath{\mathsf{last}}}
\newcommand{\mc}{\ensuremath{\mathsf{cost}}}
\newcommand{\mL}{\ensuremath{\mathsf{doc}}}
\newcommand{\mw}{\ensuremath{\mathsf{maxx}}}
\newcommand{\mi}{\ensuremath{\mathsf{maxy}}}

\newcommand{\mstainted}{\ensuremath{\texttt{Tainted}}}
\newcommand{\msset}{\ensuremath{\texttt{Set}}}

\newcommand{\pL}[4]{\ensuremath{\langle #1, #2, #3 \rangle \Downarrow_{\mathbb{RS}} #4}}
\newcommand{\pcL}[4]{\ensuremath{\langle #1, #2, #3 \rangle \Downarrow_{\mathbb{RSC}} #4}}

\newcommand{\meas}[5]{\ensuremath{\langle #1,#2,#3,\ghosted{#4},\ghosted{#5} \rangle_\mathcal{M}}}

\newcommand{\measL}[4]{\ensuremath{\langle #1, #2, #3 \rangle \Downarrow_{\mathbb{M}} #4}}

\newcommand{\msT}{\ensuremath{\mathcal{S}}}

\newcommand{\cftext}{\ensuremath{\mathsf{text}_{\mathcal{F}}}}
\newcommand{\cfnl}{\ensuremath{\mathsf{nl}_{\mathcal{F}}}}
\newcommand{\cfcombine}{\ensuremath{+_{\mathcal{F}}}}
\newcommand{\cfle}{\ensuremath{\le_{\mathcal{F}}}}

\newcommand{\cfw}{\ensuremath{\mathcal{W}}}

\newcommand{\cl}[1]{\ensuremath{\overline{#1}}}
\newcommand{\CLDocT}{\ensuremath{\cl{\mathcal{D}}_e }}
\newcommand{\render}{\ensuremath{\mathsf{eval}}}
\newcommand{\renderL}[5]{\ensuremath{\langle #1, #2, #3, #4 \rangle \Downarrow_{\mathcal{R}} #5 }}
\newcommand{\widen}[2]{\ensuremath{#1 \Downarrow_{\mathcal{W}} #2 }}
\newcommand{\obsequiv}[4]{\ensuremath{E_{#1}^{#2}(#3, #4)}}
\newcommand{\notobsequiv}[4]{\ensuremath{\neg E_{#1}^{#2}(#3, #4)}}

\newcommand{\DocS}{\ensuremath{\mathsf{Doc}}}
\newcommand{\N}{\ensuremath{\mathbb{N}}}
\newcommand{\B}{\ensuremath{\mathbb{B}}}

\newcommand{\stringconcat}{\ensuremath{+\!\!\!+\,}}

\newenvironment{proofsketch}{%
  \renewcommand{\proofname}{Proof sketch}\proof}{\endproof}

\newcommand{\memoizationLimit}{6}
\newcommand{\timeout}{\ding{36}}

\newcommand{\rulesep}{\unskip\hspace{2mm}\vrule\hspace{2mm}}

\newcommand{\separatingline}{{\color{gray}\hrule}}

\csdef{fig:wadler-non-optimality}{Figure 16}
\csdef{fig:wadler-time-complexity}{Figure 17}
\csdef{fig:bernardy-prettiness}{Figure 18}
\csdef{fig:yelland-sharing}{Figure 19}
\csdef{fig:yelland-time-complexity}{Figure 20}

\csdef{sec:appendix-survey}{Appendix A}
\csdef{sec:proofs}{Appendix B}
\csdef{sec:discussion}{Appendix C}

\newcommand{\appendixref}[1]{%
  \if\appendixmode1%
  \Cref{#1}%
  \else%
    \ifcsundef{#1}
      {\PackageError{mypackage}{appendix not handled}{extra help}}
      {\csuse{#1}}%
  \fi%
}

\begin{document}

\title{A Pretty Expressive Printer (with Appendices)}

\author{Sorawee Porncharoenwase}
\orcid{0000-0003-3900-5602}
\affiliation{%
  \department{Paul G. Allen School of Computer Science \& Engineering}
  \institution{University of Washington}
  \city{Seattle}
  \state{WA}
  \country{USA}
}
\email{sorawee@cs.washington.edu}

\author{Justin Pombrio}
\orcid{0009-0004-0244-6193}
\affiliation{%
  \institution{Unaffiliated}
  \city{Cambridge}
  \state{MA}
  \country{USA}}
\email{jpombrio@cs.brown.edu}

\author{Emina Torlak}
\orcid{0000-0002-1155-2711}
\affiliation{%
  \department{Paul G. Allen School of Computer Science \& Engineering}
  \institution{University of Washington}
  \city{Seattle}
  \state{WA}
  \country{USA}}
\email{emina@cs.washington.edu}

\begin{abstract}
Pretty printers make trade-offs between the \emph{expressiveness} of their pretty printing language, the \emph{optimality} objective that they minimize when choosing between different ways to lay out a document, and the \emph{performance} of their algorithm.
This paper presents a new pretty printer, \Pexpressive{}, that is strictly more expressive than all pretty printers in the literature and provably minimizes an optimality objective.
Furthermore, the time complexity of \Pexpressive{} is better than many existing pretty printers.
When choosing among different ways to lay out a document, \Pexpressive{} consults a user-supplied \emph{cost factory}, which determines the optimality objective, 
giving \Pexpressive{} a unique degree of flexibility.
We use the Lean theorem prover to verify the correctness (validity and optimality) of \Pexpressive{}, and
implement \Pexpressive{} concretely as a pretty printer that we call \prettiester{}.
To evaluate our pretty printer against others, we develop a formal framework for reasoning about the expressiveness of pretty printing languages, and survey pretty printers in the literature, comparing their expressiveness, optimality, worst-case time complexity, and practical running time.
Our evaluation shows that \prettiester{} is efficient and effective at producing optimal layouts. 
\prettiester{} has also seen real-world adoption: it serves as a foundation of a code formatter for Racket.
\end{abstract}



\maketitle

\def\appendixmode{1}

\section{Introduction}\label{sec:intro}

General-purpose pretty printers (or, simply, \emph{printers}) are widely used to convert structured data---typically an AST---into human-readable text. 
Their applications include code reformatting, software reengineering, and synthesized code printing~\cite{r:fmt, prettierio, reengineering, rosette:pldi}.
These printers take as inputs (1) a document in a pretty printing language (\emph{PPL}), which encodes the structured data along with formatting choices, and (2) a page width limit.
Choices in the document can yield exponentially many possible layouts.
The task of the printers then is to efficiently choose an optimal layout from all possible layouts. 
Existing printers use a variety of built-in optimality objectives.  
A good objective reflects the informal notion of ``prettiness,'' such as not overflowing past the page width limit whenever possible, 
while having as few lines as possible.

Different printers make different trade-offs in the \emph{expressiveness} of the PPL, the \emph{optimality} objective, and the \emph{performance}.
This paper presents a printer that we call \Pexpressive{}. 
It targets \Lexpressive{}, a PPL that is strictly more expressive than all published PPLs. 
This can be shown via our formal framework for reasoning about the expressiveness of PPLs. 
\Pexpressive{} is parameterized by a \emph{cost factory}, which enables users to specify an optimality objective for \Pexpressive{} to minimize.
The cost factory is versatile. 
For example, it can express non-linear costs and define concepts such as soft page width limits~\cite{yelland}.
As a result, the optimal layout that \Pexpressive{} chooses can have higher quality compared to existing printers.
The time complexity of \Pexpressive{} is $O(nW^4)$, where $n$ is the size of the document and $W$ is the computation width limit (defined in \Cref{sec:printer}).
This is better than the time complexity of many printers in the literature, and it is improved to $O(nW^3)$ when 
\Pexpressive{} is restricted to process documents in some well-known but less expressive PPLs.
We prove the correctness of \Pexpressive{} in the Lean theorem prover~\cite{lean4}, ensuring the validity and optimality of the output layout,
and demonstrate \Pexpressive{}'s efficiency by evaluating our implementation of \Pexpressive{}, which we call \prettiester{}.
We believe these attributes make \Pexpressive{} not only a good printer by itself, but also a good building block to construct other derived printers.

\paragraph{A Survey of Printers in the Wild}
To evaluate \Pexpressive{}, we conducted a broad survey of the literature on pretty printing. 
Most PPLs, embedded in a host programming language, provide a small set of core constructs that allow users to create a document with text, concatenate documents together, set indentation level, and express formatting choices.
High-level constructs can then be built on top of the core constructs.
The details of these core constructs can differ from PPL to PPL.
We found that there are two main schools of PPLs in the wild, which we call the
\emph{traditional} and \emph{arbitrary-choice} PPLs.
The traditional PPL centers around manipulation of \pp{nl}s (newlines) and current indentation level, 
while the arbitrary-choice PPL is characterized by the ability to express arbitrary formatting choices and the use of aligned concatenation to supplant the concept of indentation level.
\Cref{fig:append} illustrate documents in both PPLs that pretty-print the function definition \lstinline[language=literal]{append} in a hypothetical programming language with slightly different styling.

\begin{figure}[t]
\subcaptionbox{A document in the traditional PPL and its corresponding layouts. 
The \pp{nest} construct increments the current indentation level by some specified amount, causing \pp{nl} (newline) to insert indentation spaces. 
\pp{<>} is the unaligned concatenation operator, which places the right sub-layout after the left sub-layout on the current indentation level.
Lastly, the \pp{group} construct creates a choice between two alternatives: one where the sub-layouts are left alone and one where the sub-layouts are flattened by replacing newlines and indentation spaces due to \pp{nl}s in the group with single spaces.
\label{fig:append:traditional}}[\hsize]{
  \parbox{\figrasterwd}{
    \parbox{.52\figrasterwd}{%
      \begin{lstlisting}[language=pp]
text "function append(first,second,third){"(&|&)
<> nest 4 ((&|&)
  let f = text "first +" in(&|&)
  let s = text "second +" in(&|&)
  let t = text "third" in(&|&)
  nl <> (!\tikzmark{gappend-traditional-code1l}!)text "return "(!\tikzmark{gappend-traditional-code2r}!) <>(&|&)
  (!\tikzmark{gappend-traditional-code3l}!)group (nest 4 (f <> nl <> s <> nl <> t))(!\tikzmark{gappend-traditional-code4r}!)(&|&)
) <> nl <> text "}"(&|&)
\end{lstlisting}
\begin{tikzpicture}[remember picture,overlay]
\fill[rounded corners=1pt, fill=blue, fill opacity=0.15]
([shift={(0pt,-2pt)}]pic cs:gappend-traditional-code4r)
--
([shift={(0pt,-2pt)}]pic cs:gappend-traditional-code3l)
--
([shift={(0pt,7pt)}]pic cs:gappend-traditional-code3l)
--
([shift={(0pt,7pt)}]pic cs:gappend-traditional-code4r)
--
([shift={(0pt,-2pt)}]pic cs:gappend-traditional-code4r);
\fill[rounded corners=1pt, fill=red, fill opacity=0.2]
([shift={(0pt,7pt)}]pic cs:gappend-traditional-code1l)
--
([shift={(0pt,7pt)}]pic cs:gappend-traditional-code2r)
--
([shift={(0pt,-2pt)}]pic cs:gappend-traditional-code2r)
--
([shift={(0pt,-2pt)}]pic cs:gappend-traditional-code1l)
--
([shift={(0pt,7pt)}]pic cs:gappend-traditional-code1l);
\end{tikzpicture}
      \vspace{-4mm} 
    }\hskip2em
    \parbox{.46\figrasterwd}{%
      \begin{lstlisting}[numbers=left, numbersep=.75em, language=literal]
function append(first,(!\tikzmark{gappend-traditional-vert1l}!)second,third){(!\tikzmark{gappend-traditional-vert2l}!) (&|&)
    (!\tikzmark{gappend-traditional-vert4l}!)return (!\tikzmark{gappend-traditional-vert5r}!)(!\tikzmark{gappend-traditional-vert6l}!)first +(!\tikzmark{gappend-traditional-vert7r}!)(&|&)
    (!\tikzmark{gappend-traditional-vert9l}!)    sec(!\tikzmark{gappend-traditional-vert11l}!)on(!\tikzmark{gappend-traditional-vert12r}!)d +  (!\tikzmark{gappend-traditional-vert13r}!)(&|&)
    (!\tikzmark{gappend-traditional-vert15l}!)    third(!\tikzmark{gappend-traditional-vert17r}!)(&|&)
}                     (!\tikzmark{gappend-traditional-vert18l}!)              (!\tikzmark{gappend-traditional-vert19l}!) (&|&)
\end{lstlisting}
\begin{tikzpicture}[remember picture,overlay]
\fill[rounded corners=1pt, fill=blue, fill opacity=0.15]
([shift={(0pt,7pt)}]pic cs:gappend-traditional-vert7r)
--
([shift={(0pt,-2pt)}]pic cs:gappend-traditional-vert13r)
--
([shift={(0pt,-2pt)}]pic cs:gappend-traditional-vert12r)
--
([shift={(0pt,-2pt)}]pic cs:gappend-traditional-vert17r)
--
([shift={(0pt,-2pt)}]pic cs:gappend-traditional-vert15l)
--
([shift={(0pt,7pt)}]pic cs:gappend-traditional-vert9l)
--
([shift={(0pt,7pt)}]pic cs:gappend-traditional-vert11l)
--
([shift={(0pt,7pt)}]pic cs:gappend-traditional-vert6l)
--
([shift={(0pt,7pt)}]pic cs:gappend-traditional-vert7r);
\fill[rounded corners=1pt, fill=red, fill opacity=0.2]
([shift={(0pt,7pt)}]pic cs:gappend-traditional-vert4l)
--
([shift={(0pt,7pt)}]pic cs:gappend-traditional-vert5r)
--
([shift={(0pt,-2pt)}]pic cs:gappend-traditional-vert5r)
--
([shift={(0pt,-2pt)}]pic cs:gappend-traditional-vert4l)
--
([shift={(0pt,7pt)}]pic cs:gappend-traditional-vert4l);
\draw[red, ultra thick, densely dashed]
([shift={(0pt,7pt)}]pic cs:gappend-traditional-vert1l)
--
([shift={(0pt,-2pt)}]pic cs:gappend-traditional-vert18l);
\draw[teal, ultra thick, densely dashed]
([shift={(0pt,7pt)}]pic cs:gappend-traditional-vert2l)
--
([shift={(0pt,-2pt)}]pic cs:gappend-traditional-vert19l);
\end{tikzpicture}
      \vspace{-3mm} 
      \begin{lstlisting}[numbers=left, numbersep=.75em, language=literal]
function append(first,(!\tikzmark{gappend-traditional-horz1l}!)second,third){(!\tikzmark{gappend-traditional-horz2l}!) (&|&)
    (!\tikzmark{gappend-traditional-horz4l}!)return (!\tikzmark{gappend-traditional-horz5r}!)(!\tikzmark{gappend-traditional-horz6l}!)first + second + third(!\tikzmark{gappend-traditional-horz7r}!)(&|&)
}                     (!\tikzmark{gappend-traditional-horz8l}!)              (!\tikzmark{gappend-traditional-horz9l}!) (&|&)
\end{lstlisting}
\begin{tikzpicture}[remember picture,overlay]
\fill[rounded corners=1pt, fill=blue, fill opacity=0.15]
([shift={(0pt,7pt)}]pic cs:gappend-traditional-horz7r)
--
([shift={(0pt,-2pt)}]pic cs:gappend-traditional-horz7r)
--
([shift={(0pt,-2pt)}]pic cs:gappend-traditional-horz6l)
--
([shift={(0pt,7pt)}]pic cs:gappend-traditional-horz6l)
--
([shift={(0pt,7pt)}]pic cs:gappend-traditional-horz7r);
\fill[rounded corners=1pt, fill=red, fill opacity=0.2]
([shift={(0pt,7pt)}]pic cs:gappend-traditional-horz4l)
--
([shift={(0pt,7pt)}]pic cs:gappend-traditional-horz5r)
--
([shift={(0pt,-2pt)}]pic cs:gappend-traditional-horz5r)
--
([shift={(0pt,-2pt)}]pic cs:gappend-traditional-horz4l)
--
([shift={(0pt,7pt)}]pic cs:gappend-traditional-horz4l);
\draw[red, ultra thick, densely dashed]
([shift={(0pt,7pt)}]pic cs:gappend-traditional-horz1l)
--
([shift={(0pt,-2pt)}]pic cs:gappend-traditional-horz8l);
\draw[teal, ultra thick, densely dashed]
([shift={(0pt,7pt)}]pic cs:gappend-traditional-horz2l)
--
([shift={(0pt,-2pt)}]pic cs:gappend-traditional-horz9l);
\end{tikzpicture}
      \vspace{-4mm} 
    }
  }
}
\subcaptionbox{A document in the arbitrary-choice PPL and its corresponding layouts.
\pp{<|>} is the arbitrary-choice operator, which per its namesake, creates a choice between the layouts of two arbitrary sub-documents.
\pp{<$>} is the vertical concatenation operator, which joins two sub-layouts with a newline.
Lastly, \pp{<+>} is the aligned concatenation operator, which joins two sub-layouts horizontally, aligning the whole right sub-layout at the column where it is to be placed in.\label{fig:append:arbitrary}}[\hsize]{
  \parbox{\figrasterwd}{
    \parbox{.52\figrasterwd}{%
      \begin{lstlisting}[language=pp]
text "function append(first,second,third){" <$>(&|&)
( let f = text "first +" in(&|&)
  let s = text "second +" in(&|&)
  let t = text "third" in(&|&)
  let sp = text " " in(&|&)
  let ret = text "return " in(&|&)
  (!\tikzmark{gappend-arbitrary-code1l}!)text "    "(!\tikzmark{gappend-arbitrary-code2r}!) <+>(&|&)
  ((!\tikzmark{gappend-arbitrary-code3l}!)((ret <+> text "(") <$>               (!\tikzmark{gappend-arbitrary-code4r}!)(&|&)
    (text  (!\tikzmark{gappend-arbitrary-code5r}!)  (!\tikzmark{gappend-arbitrary-code6r}!)    <+> (f <$> s <$> t)) <$>(!\tikzmark{gappend-arbitrary-code7r}!)(&|&)
   (!\tikzmark{gappend-arbitrary-code8l}!) text ")")(!\tikzmark{gappend-arbitrary-code9r}!) <|>(&|&)
   (!\tikzmark{gappend-arbitrary-code10l}!)(ret <+> f <+> sp <+> s <+> sp <+> t)(!\tikzmark{gappend-arbitrary-code11r}!)))(&|&)
<$> text "}"(&|&)
\end{lstlisting}
\begin{tikzpicture}[remember picture,overlay]
\fill[rounded corners=1pt, fill=blue, fill opacity=0.15]
([shift={(0pt,7pt)}]pic cs:gappend-arbitrary-code4r)
--
([shift={(0pt,-2pt)}]pic cs:gappend-arbitrary-code7r)
--
([shift={(0pt,-2pt)}]pic cs:gappend-arbitrary-code6r)
--
([shift={(0pt,-2pt)}]pic cs:gappend-arbitrary-code9r)
--
([shift={(0pt,-2pt)}]pic cs:gappend-arbitrary-code8l)
--
([shift={(0pt,7pt)}]pic cs:gappend-arbitrary-code3l)
--
([shift={(0pt,7pt)}]pic cs:gappend-arbitrary-code4r);
\fill[rounded corners=1pt, fill=yellow, fill opacity=0.2]
([shift={(0pt,7pt)}]pic cs:gappend-arbitrary-code10l)
--
([shift={(0pt,7pt)}]pic cs:gappend-arbitrary-code11r)
--
([shift={(0pt,-2pt)}]pic cs:gappend-arbitrary-code11r)
--
([shift={(0pt,-2pt)}]pic cs:gappend-arbitrary-code10l)
--
([shift={(0pt,7pt)}]pic cs:gappend-arbitrary-code10l);
\fill[rounded corners=1pt, fill=red, fill opacity=0.2]
([shift={(0pt,7pt)}]pic cs:gappend-arbitrary-code1l)
--
([shift={(0pt,7pt)}]pic cs:gappend-arbitrary-code2r)
--
([shift={(0pt,-2pt)}]pic cs:gappend-arbitrary-code2r)
--
([shift={(0pt,-2pt)}]pic cs:gappend-arbitrary-code1l)
--
([shift={(0pt,7pt)}]pic cs:gappend-arbitrary-code1l);
\draw
([shift={(0pt,2pt)}]pic cs:gappend-arbitrary-code5r)node[]{$\quad\ \ \ $\lstinline[language=literal, showspaces]!"    "!};
\end{tikzpicture}
      \vspace{-4mm} 
    }\hskip2em
    \parbox{.46\figrasterwd}{%
      \begin{lstlisting}[numbers=left, numbersep=.75em, language=literal]
function append(first,(!\tikzmark{gappend-arbitrary-vert1l}!)second,third){(!\tikzmark{gappend-arbitrary-vert2l}!) (&|&)
(!\tikzmark{gappend-arbitrary-vert3l}!)    (!\tikzmark{gappend-arbitrary-vert4r}!)(!\tikzmark{gappend-arbitrary-vert5l}!)return (    (!\tikzmark{gappend-arbitrary-vert6r}!)(&|&)
        first +(&|&)
        second +(&|&)
     (!\tikzmark{gappend-arbitrary-vert7r}!)   third   (!\tikzmark{gappend-arbitrary-vert8r}!)(&|&)
    (!\tikzmark{gappend-arbitrary-vert9l}!))(!\tikzmark{gappend-arbitrary-vert9r}!)(&|&)
}                     (!\tikzmark{gappend-arbitrary-vert10l}!)              (!\tikzmark{gappend-arbitrary-vert11l}!) (&|&)
\end{lstlisting}
\begin{tikzpicture}[remember picture,overlay]
\fill[rounded corners=1pt, fill=blue, fill opacity=0.15]
([shift={(0pt,7pt)}]pic cs:gappend-arbitrary-vert6r)
--
([shift={(0pt,-2pt)}]pic cs:gappend-arbitrary-vert8r)
--
([shift={(0pt,-2pt)}]pic cs:gappend-arbitrary-vert7r)
--
([shift={(0pt,-2pt)}]pic cs:gappend-arbitrary-vert9r)
--
([shift={(0pt,-2pt)}]pic cs:gappend-arbitrary-vert9l)
--
([shift={(0pt,7pt)}]pic cs:gappend-arbitrary-vert5l)
--
([shift={(0pt,7pt)}]pic cs:gappend-arbitrary-vert6r);
\fill[rounded corners=1pt, fill=red, fill opacity=0.2]
([shift={(0pt,7pt)}]pic cs:gappend-arbitrary-vert3l)
--
([shift={(0pt,7pt)}]pic cs:gappend-arbitrary-vert4r)
--
([shift={(0pt,-2pt)}]pic cs:gappend-arbitrary-vert4r)
--
([shift={(0pt,-2pt)}]pic cs:gappend-arbitrary-vert3l)
--
([shift={(0pt,7pt)}]pic cs:gappend-arbitrary-vert3l);
\draw[red, ultra thick, densely dashed]
([shift={(0pt,7pt)}]pic cs:gappend-arbitrary-vert1l)
--
([shift={(0pt,-2pt)}]pic cs:gappend-arbitrary-vert10l);
\draw[teal, ultra thick, densely dashed]
([shift={(0pt,7pt)}]pic cs:gappend-arbitrary-vert2l)
--
([shift={(0pt,-2pt)}]pic cs:gappend-arbitrary-vert11l);
\end{tikzpicture}
      \vspace{-2mm} 
      \begin{lstlisting}[numbers=left, numbersep=.75em, language=literal]
function append(first,(!\tikzmark{gappend-arbitrary-horz1l}!)second,third){(!\tikzmark{gappend-arbitrary-horz2l}!) (&|&)
(!\tikzmark{gappend-arbitrary-horz3l}!)    (!\tikzmark{gappend-arbitrary-horz4r}!)(!\tikzmark{gappend-arbitrary-horz5l}!)return first + second + third(!\tikzmark{gappend-arbitrary-horz6r}!)(&|&)
}                     (!\tikzmark{gappend-arbitrary-horz7l}!)              (!\tikzmark{gappend-arbitrary-horz8l}!) (&|&)
\end{lstlisting}
\begin{tikzpicture}[remember picture,overlay]
\fill[rounded corners=1pt, fill=yellow, fill opacity=0.2]
([shift={(0pt,7pt)}]pic cs:gappend-arbitrary-horz6r)
--
([shift={(0pt,-2pt)}]pic cs:gappend-arbitrary-horz6r)
--
([shift={(0pt,-2pt)}]pic cs:gappend-arbitrary-horz5l)
--
([shift={(0pt,7pt)}]pic cs:gappend-arbitrary-horz5l)
--
([shift={(0pt,7pt)}]pic cs:gappend-arbitrary-horz6r);
\fill[rounded corners=1pt, fill=red, fill opacity=0.2]
([shift={(0pt,7pt)}]pic cs:gappend-arbitrary-horz3l)
--
([shift={(0pt,7pt)}]pic cs:gappend-arbitrary-horz4r)
--
([shift={(0pt,-2pt)}]pic cs:gappend-arbitrary-horz4r)
--
([shift={(0pt,-2pt)}]pic cs:gappend-arbitrary-horz3l)
--
([shift={(0pt,7pt)}]pic cs:gappend-arbitrary-horz3l);
\draw[red, ultra thick, densely dashed]
([shift={(0pt,7pt)}]pic cs:gappend-arbitrary-horz1l)
--
([shift={(0pt,-2pt)}]pic cs:gappend-arbitrary-horz7l);
\draw[teal, ultra thick, densely dashed]
([shift={(0pt,7pt)}]pic cs:gappend-arbitrary-horz2l)
--
([shift={(0pt,-2pt)}]pic cs:gappend-arbitrary-horz8l);
\end{tikzpicture}
      \vspace{-4mm} 
    }
  }    
}
\caption{
The traditional and arbitrary-choice PPLs, embedded in the host language OCaml.
Colored regions in a document and corresponding layouts indicate the correspondence between the colored sub-documents and the colored sub-layouts.
We use the \pp{let} construct to make the documents easier to read, even though it is usually not a part of PPLs.
Dotted lines illustrate different page width limits at 22 and 36 characters.
}
\label{fig:append}
\end{figure}

\paragraph{Expressiveness} The literature contains informal claims about the expressiveness of PPLs~\cite{wadler, chitil, podkopaev}.
We develop two formal notions of expressiveness: the ability to \emph{express layouts} and the ability to \emph{express features}.
The former reflects the functionality of a PPL, while the latter reflects the ease of document construction.
Using our framework, we can show that neither the traditional PPL nor the arbitrary-choice PPL is more expressive than the other. 
For example, the set of layouts in \Cref{fig:append:arbitrary} cannot be expressed by any document in the traditional PPL. 
This is because all layouts due to a particular document in the traditional PPL must be the same modulo whitespace, but one of the layouts in the figure has an extra pair of parentheses.\footnote{
  Languages such as Python require an extra pair of parentheses around an expression that spans multiple lines~\cite{implicit-line-join}. 
  Similarly, some styles prefer adding an extra comma (also known as trailing comma) when a function call spans multiple lines~\cite{trailing-comma}.
  Hence, the ability to express layouts with differing content is desirable.
}
As another example, the document in \Cref{fig:append:arbitrary} is awkwardly constructed, because the document structure and the underlying AST structure do not match (\Cref{sec:framework:definability}).
It would be more natural to use unaligned concatenation, but the feature cannot be expressed by any combination of features in the arbitrary-choice PPL.\footnote{
  Different programming language styles prefer different concatenation operators.
  C-like languages heavily use unaligned concatenation, while aligned concatenation has been used for Haskell, Lisp, R, and Julia. 
  However, there are instances where C-like languages would benefit from aligned concatenation, and Haskell would benefit from unaligned concatenation.
}
To that end, we develop a PPL called \Lexpressive{} that is strictly more expressive than both the traditional and arbitrary-choice PPLs, facilitating both functionality and ease of document construction.

\paragraph{Optimality} The optimality objective of a printer indicates what it optimizes for when resolving choices.
Most printers targeting the traditional PPL minimize overflow past the page width limit line-by-line, preferring a longer line when there is no overflow.
For example, given the document in \Cref{fig:append:traditional}, the first layout is optimal when the page width limit is 22 (red dotted line), while the second layout is optimal when the page width limit is 36 (green dotted line).
Contrary to prior claims~\cite{wadler,chitil}, we discovered that this strategy guarantees neither the absence of overflow whenever possible nor the minimality of the number of lines.
By contrast, most printers targeting the arbitrary-choice PPL minimize the number of lines among layouts with no overflow.
However, they \emph{error} when all possible layouts have an overflow, resulting in a poor user experience (e.g., when the page width limit is 22 in \Cref{fig:append:arbitrary}).
Recognizing that unavoidable overflows do occur in practice, we introduce the concept of a \emph{cost factory}, which allows users to choose a desired objective permitted by its interface, including an objective that tolerates overflow gracefully.

\paragraph{Performance} 
Printing proceeds in two phases: resolving choices and rendering the optimal choice to text (although many printers fuse these two phases together).
Time complexity of printers is best measured against the resolving phase\footnote{This formulation allows us to talk about ``linear-time'' printers, even though there are, e.g., documents whose size is $O(n)$, but its optimal layout has $O(n^2)$ characters.}, and it is usually specified with two parameters: the size of the document $n$ and the width limit $W$, with the preference that the time complexity be polynomial in $W$ and linear in $n$.
Most printers in the literature leave their time complexity unanalyzed, instead opting to show experimental results that their implementations are efficient in practice.
We analyze these printers and demonstrate documents that trigger worse than linear time behavior (in $n$) on some printers. 
Further complication arises in printers with the arbitrary choice feature, which gives rise to documents that are structured as DAGs as opposed to trees.
We show that many printers that treat the input document as a tree suffer from a combinatorial explosion as the DAG structure is unfolded during the resolving phase, resulting in exponential time complexity.
With a combination of proof and experimental results, we show that the time complexity of \Pexpressive{} is linear in the DAG size of the document and that it runs fast in practice. \\

In summary, this paper makes the following contributions:

\begin{itemize}
  \item A new PPL called \Lexpressive{} that is strictly more expressive than all published PPLs.
  The constructs in \Lexpressive{} are not new, but packaging them all in a single PPL has never been done before.
  \item A printer \Pexpressive{} targeting \Lexpressive{} that utilizes a \emph{cost factory} to allow a variety of optimality objectives.
  \item A proof of correctness (validity and optimality) for \Pexpressive{}, formalized in the Lean theorem prover.
  To our knowledge, this is the first time that a printer has been formally verified.
  \item A framework to formally reason about the expressiveness of PPLs.
  \item A survey of printers and an analysis that dispels common misunderstandings about them.
  \item An implementation of \Pexpressive{}, \prettiester, and an evaluation that shows its effectiveness.
\end{itemize}

The rest of this paper is structured as follows. 
\Cref{sec:related} surveys the related work.
\Cref{sec:tour} provides an overview of \Pexpressive{} from the user's perspective.
\Cref{sec:semantics} presents the formal semantics of \Lexpressive{}.
\Cref{sec:framework} introduces a framework to reason about the expressiveness of PPLs.
\Cref{sec:printer} formally presents \Pexpressive{} and its analysis.
\Cref{sec:impl} discusses \prettiester{}, an implementation of \Pexpressive{}.
\Cref{sec:eval} presents an evaluation of \prettiester{} that demonstrates its effectiveness.
Lastly, \Cref{sec:conclusion} concludes the paper.

\section{Related work}\label{sec:related}

To understand the trade-off space of printer designs, we conduct a comprehensive analysis of related work in the literature.
This section provides our analysis of the printers, grouped by the expressiveness of their public interface\footnote{In practice, printers include extensions that increase their expressiveness. 
A printer may even have different expressiveness across different versions.
This section focuses on the core features of these printers as specified in their publications.}.
The summary is presented in \Cref{table:comparison}.
We then compare and contrast our printer \Pexpressive{} against them.

\begin{table}[t]
\scriptsize\centering
\caption{A comparison of existing printers. 
$n$ and $\hat{n}$ are the DAG size and tree size of the input document (where $\hat{n}$ in the worst case is exponential in $n$).
$W$ is the width limit.
}
\label{table:comparison}
\begin{threeparttable}
\begin{tabular}{lllll}
\toprule
  & \multicolumn{2}{c}{\textbf{Expressiveness}}
  & \textbf{Optimality}
  & \textbf{Performance}
\\
  \cmidrule(r){2-3}
  \cmidrule(r){4-4}
  \cmidrule(r){5-5}

\textbf{Printer}
                  & Choice                 & Concatenation
                  & Minimization objective
                  & Time complexity \\
\midrule
\citet{oppen}     & Group                  & Unaligned
                  & Lexicographic overflow
                  & $O(n)$\\
\citet{hughes}    & Group                  & Aligned
                  & Lexicographic overflow
                  & $O(n^2)$ \\
\citet{wadler}    & Group                  & Unaligned
                  & Lexicographic overflow
                  & $O(n^2)$ \\
\citet{leijen}    & Group                  & Both
                  & Lexicographic overflow
                  & $O(n^2)$ \\
\citet{chitil}    & Group                  & Unaligned
                  & Lexicographic overflow
                  & $O(n)$ \\
\citet{kiselyov}  & Group                  & Unaligned
                  & Lexicographic overflow
                  & $O(n)$ \\
\citet{swierstra} & Arbitrary              & Aligned
                  & Height$^\dagger$
                  & Exp. in $n$ \\  
\citet{podkopaev} & Arbitrary              & Aligned
                  & Height$^\dagger$
                  & $O(\hat{n} W^4)$ \\
\citet{yelland}   & Arbitrary              & Aligned
                  & Linear cost
                  & $O(\hat{n}^{3/2})$ \\
\citet{bernardy}  & Arbitrary              & Aligned
                  & Height$^\dagger$
                  & $O(n W^6)$ \\
\midrule
\Pexpressive{}    & Both              & Both
                  & Cost (from the cost factory)
                  & $O(nW^4)$ \\
\Pexpressive{} (aligned only)
                  & Both              & Aligned
                  & Cost (from the cost factory)
                  & $O(nW^3)$ \\
\bottomrule
\end{tabular}

\begin{tablenotes}
\footnotesize
\item[$\dagger$] only consider layouts without an overflow past $W$.
\end{tablenotes}
\end{threeparttable}
\end{table}

\subsection{Traditional Printers}

Pretty printing has a long history.
\citet{oppen} first introduced a general-purpose printer, written in the imperative style.
Oppen pioneered the PPL that we call the traditional PPL, shown in \Cref{fig:grammar:traditional}.
Instead of representing an input document as a tree, as commonly done in subsequent work, Oppen represents the document as a stream of ``instruction tokens.''
The algorithm's time complexity is $O(n)$, where $n$ is the length of the stream.
Furthermore, the algorithm is \emph{bounded}, requiring a limited look-ahead into the stream.
As with other printers in this family, the printer greedily minimizes overflow past the page width limit, which neither avoids overflow whenever possible nor minimizes number of lines, as discussed in Oppen's paper.

\begin{figure}[t]
\begin{subfigure}[t]{.53\textwidth}
    \footnotesize
\begin{tabular}{r@{\hspace{.5em}}lll}
$d \in \mathcal{D}\ \defn$ &
\lstinline[language=pp]!text! $s$ & text \\ 
$\vbar$ & \lstinline[language=pp]!nl! & newline \\
$\vbar$ & $d\ $ \lstinline[language=pp]!<>! $d$ & unaligned concatenation \\ 
$\vbar$ & \lstinline[language=pp]!nest! $ n\ d$ & increase the indentation level by $n$ \\ 
$\vbar$ & \lstinline[language=pp]!group! $d$ & a choice between \\ 
& & flattening or not flattening \\
\end{tabular}%

    \caption{A variant of traditional PPL from \citet{wadler}.}
    \label{fig:grammar:traditional}
\end{subfigure}\hspace{1mm}%
\begin{subfigure}[t]{.46\textwidth}
    \footnotesize
\begin{tabular}{r@{\hspace{.5em}}ll}
$d \in \mathcal{D}\ \defn$ &
\lstinline[language=pp]!text! $s$ & text \\
$\vbar$ & $d_a$ \lstinline[language=pp]!<+>! $d_b$ & aligned concatenation \\
$\vbar$ & $d_a$ \lstinline[language=pp]!<$>! $d_b$ & vertical concatenation \\ 
$\vbar$ & $d_a$ \lstinline[language=pp]!<|>! $d_b$ & an arbitrary choice \\ 
\end{tabular}%

    \caption{A variant of arbitrary-choice PPL from \\ \citet{podkopaev}.}
    \label{fig:grammar:arbitrary}
\end{subfigure}
\label{fig:grammar}
\caption{A comparison between the traditional and arbitrary-choice PPLs.
$s$ denotes a string without newline, and $n$ denotes a natural number.}
\end{figure}

\citet{wadler} designed a printer that targets the traditional PPL. 
It is used in many real world applications, such as an industrial code formatter~\cite{prettierio}, and as a basis for much pretty printing research~\cite{chitil,kiselyov}.
The printer aims to be a rewrite of Oppen's printer using the functional style employed by Hughes (described later).
The printer is claimed~\cite{wadler,chitil} to produce an output layout that does not exceed the width limit whenever possible, and minimizes the number of lines. 
However, this is not the case, as shown in \appendixref{fig:wadler-non-optimality} in \appendixref{sec:appendix-survey}.
The time complexity of the printer is claimed to be $O(n)$ where $n$ is the size of document~\cite{wadler}, but it is in fact $O(n^2)$ in the worst case, as demonstrated in \appendixref{fig:wadler-time-complexity}, although this worst case behavior is unlikely to occur in practice.

\citet{chitil} improved Wadler's printer so that it is as efficient as Oppen's, $O(n)$, by using lazy dequeues.
\citet{kiselyov} similarly improved  Wadler's printer via their generator framework.

Compared to traditional printers, \Pexpressive{} is more expressive as it allows arbitrary choices and aligned concatenation. 
Furthermore, \Pexpressive{} can produce an output layout that minimizes number of lines when the output layout does not exceed the page width limit, and does not exceed the page width limit whenever possible.
The tradeoff is that \Pexpressive{} is less space efficient and slower than traditional printers. 
The space complexity of traditional printers is sub-linear in the size of document, which was especially important decades ago when memory is scarce. The space complexity of \Pexpressive{}  is $O(nW^3)$ in the worst case (or $O(nW^2)$ when targeting some PPLs).  We find that, on modern machines, the added memory consumption and performance overhead are rarely an issue in practice (\Cref{sec:eval}).

\subsection{Arbitrary-Choice Printers}\label{sec:related:arbitrary}

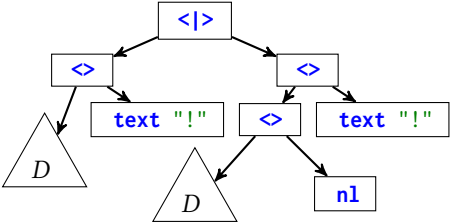
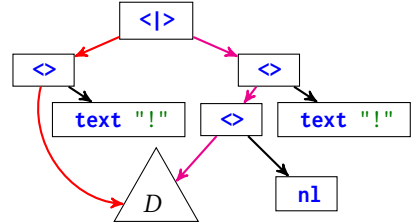
\begin{figure}[t]
\centering
\begin{subfigure}[t]{.9\textwidth}
\begin{lstlisting}[language=pp]
let shared := (!$D$!) in (shared <> text "!") <|> ((shared <> nl) <> text "!")
\end{lstlisting}
    \caption{
    A document that encodes (at least) two possible layouts.
    $D$ is an arbitrary sub-document.
    }
    \label{fig:sharing:code}
\end{subfigure}\vskip .5em
\begin{subfigure}[t]{.45\textwidth}
    \centering
    \begin{tikzpicture}[align=left]
  \newcommand\sespace{-0.65} 
  \tikzstyle{state} = [draw, rectangle, font=\linespread{.8}\selectfont];
  \tikzstyle{tree} = [draw, isosceles triangle, shape border rotate=90, isosceles triangle apex angle=60, font=\linespread{.8}\selectfont];
  \tikzstyle{edge} = [draw, thick, ->, >=stealth'];
  \tikzstyle{location} = [draw,circle,color=orange!70!black,inner sep=1pt,minimum width=1em]
  
\begin{scope}[shift={(0,0)}]
  \node[state,name=s0] at (0, 0) {
    \enspace\pp{<|>}
  };
  
  \node[state,name=s1] at ($(-1.5, 1*\sespace)$) {
    \enspace\pp{<>}
  };
  
  \node[state,name=s2] at ($(+1.5, 1*\sespace)$) {
    \enspace\pp{<>}
  };

  \node[state,name=s3] at ($(-0.5, 2*\sespace)$) {
    \enspace\pp{text "!"}
  };
    
  \node[state,name=s4] at ($(+2.5, 2*\sespace)$) {
    \enspace\pp{text "!"}
  };
  
  \node[tree,name=shared1] at ($(-2, 3*\sespace)$) {
    $D$
  };

  \node[state,name=s5] at ($(1, 2*\sespace)$) {
    \enspace\pp{<>}
  };

  \node[state,name=s6] at ($(2, 3.5*\sespace)$) {
    \enspace\pp{nl}
  };

  \node[tree,name=shared2] at ($(0, 3.7*\sespace)$) {
    $D$
  };
  
  \path[edge] (s0) -- (s1);
  \path[edge] (s0) -- (s2);
  \path[edge] (s1) -- (shared1);
  \path[edge] (s2) -- (s5);
  \path[edge] (s2) -- (s4);
  \path[edge] (s5) -- (shared2);
  \path[edge] (s5) -- (s6);
  \path[edge] (s1) -- (s3);
  
  \end{scope}
 
\end{tikzpicture}
    \caption{A tree representation of \Cref{fig:sharing:code}. \\
    $D$ contributes to the size twice.}
    \label{fig:sharing:tree}
\end{subfigure}\hspace{5mm}%
\begin{subfigure}[t]{.45\textwidth}
    \centering
    \begin{tikzpicture}[align=left]
  \newcommand\sespace{-0.65} 
  \tikzstyle{state} = [draw, rectangle, font=\linespread{.8}\selectfont];
  \tikzstyle{tree} = [draw, isosceles triangle, shape border rotate=90, isosceles triangle apex angle=60, 
  font=\linespread{.8}\selectfont];
  \tikzstyle{edge} = [draw, thick, ->, >=stealth'];
  \tikzstyle{location} = [draw,circle,color=orange!70!black,inner sep=1pt,minimum width=1em]

\begin{scope}[shift={(0,0)}]
  \node[state,name=s0] at (0, 0) {
    \enspace\pp{<|>}
  };
  
  \node[state,name=s1] at ($(-1.5, 1*\sespace)$) {
    \enspace\pp{<>}
  };
  
  \node[state,name=s2] at ($(+1.5, 1*\sespace)$) {
    \enspace\pp{<>}
  };

  \node[state,name=s3] at ($(-0.5, 2*\sespace)$) {
    \enspace\pp{text "!"}
  };
    
  \node[state,name=s4] at ($(+2.5, 2*\sespace)$) {
    \enspace\pp{text "!"}
  };

  \node[state,name=s5] at ($(1, 2*\sespace)$) {
    \enspace\pp{<>}
  };

  \node[state,name=s6] at ($(2, 3.5*\sespace)$) {
    \enspace\pp{nl}
  };

  \node[tree,name=shared] at ($(0, 3.7*\sespace)$) {
    $D$
  };
  
  \path[edge] (s0) [color=red] -- (s1);
  \path[edge] (s0) [color=magenta] -- (s2);
  \path[edge] (s1) edge[bend right=50] [color=red] (shared);
  \path[edge] (s2) [color=magenta] -- (s5);
  \path[edge] (s2) -- (s4);
  \path[edge] (s5) [color=magenta] -- (shared);
  \path[edge] (s5) -- (s6);
  \path[edge] (s1) -- (s3);
  
  \end{scope}
 
\end{tikzpicture}
    \caption{A DAG representation of \Cref{fig:sharing:code}.\\
    $D$ contributes to the size only once.}
    \label{fig:sharing:dag}
\end{subfigure}
\caption{An example document that shows the importance of treating document as a DAG rather than a tree. 
The red and pink paths illustrate that the DAG is \emph{properly shared}, as will be discussed in \Cref{sec:printer:doc-structure}.}
\label{fig:sharing}
\end{figure}

\citet{azero} introduced a printer that supports aligned concatenation and choices between arbitrary alternatives.
It started the line of work that targets the arbitrary-choice PPL, shown in \Cref{fig:grammar:arbitrary}.
The printer's optimality objective is to avoid overflow whenever possible and produce a minimal number of lines. 
However, it does not have the ability to cope with unavoidable overflow. 
This printer was soon superseded by \citet{swierstra}, which improves its performance via heuristics and adds the capability to \emph{share} a sub-document across choices by deeply embedding the (equivalent of a) \pp{let} construct in the PPL.
As a result, the later printer can process documents that are structured as DAGs rather than trees, as shown in \Cref{fig:sharing}.
Nonetheless, the time complexity of both printers is exponential in $n$~\cite{podkopaev}.\tighten

\citet{podkopaev} improved upon Swierstra et al.'s work by formulating the problem as  dynamic programming. 
This fixes the exponential blowup in the prior work, but treats the document as a tree, making its time complexity $O(\hat{n}W^4)$, where $\hat{n}$ is the tree size of the document, which could be exponentially larger than its DAG size. 
The paper acknowledges the problem and surmises that memoization may be able to address it.

The paper by \citet{bernardy} is the main inspiration for our work. 
The printer uses Pareto frontiers to find an optimal layout.
By shallowly embedding the PPL (in Haskell), computations on sub-documents are effectively shared for free.
However, as presented in the paper, the printer requires the page width limit to be hard-coded.
In the actual implementation~\cite{bernardy:implementation}, the page width limit is customizable, accomplished by threading the value through functions.
But this change destroys the shared computations, leading to exponential running time.
Compared to \citet{podkopaev}'s work, \citet{bernardy}'s approach can exploit sparseness to improve practical efficiency, 
but the use of an inefficient algorithm makes the time complexity of the printer $O(nW^6)$ in the worst case.
While the paper does not handle unavoidable overflow,
the implementation does by automatically scaling up the page width limit (or equivalently, minimizing the maximum overflow). 
This, however, allows avoidable overflow elsewhere, as shown in \appendixref{fig:bernardy-prettiness} in \appendixref{sec:appendix-survey}, which is undesirable.
Later on, Bernardy abandoned the arbitrary-choice operator, noting that it could trigger the exponential behavior~\cite{bernardy:disjunctionless}.

\citet{yelland} similarly targeted the arbitrary-choice PPL.
However, the paper took a very different approach. 
The core printer restricts the use of aligned concatenation by requiring the left sub-document to be a \pp{text} syntactically.
This restriction allows the core printer to utilize the concept of ``piecewise linear cost function'' to seemingly boost its performance.
To achieve the expressiveness of the arbitrary-choice PPL, the printer employs rewriting rules to transform the original document into the restricted document.
While the work carefully avoids exponential blowup by sharing sub-documents in the resulting restricted document,
it does not necessarily preserve the sharing structure of the original document, as demonstrated in \appendixref{fig:yelland-sharing} in \appendixref{sec:appendix-survey}. 
Compound this with the lack of a computation width limit, and the number of piecewise linear cost functions under consideration could be as large as $O(\hat{n}^{1/2})$, making the time complexity $O(\hat{n}^{3/2})$ in total, as shown in \appendixref{fig:yelland-time-complexity}.
Another aspect to consider is the printer's optimality objective, which is restricted to minimizing a linear combination of quantities like the number of lines and overflow. 
Hence, the work will not technically avoid overflow whenever possible (although the overflow coefficient can be made arbitrarily large to arbitrarily discourage overflow).
On the other hand, this optimality objective can support unique features, such as incorporating the costs due to multiple soft page width limits.

Compared to arbitrary-choice printers, \Pexpressive{} is more expressive as it allows unaligned concatenation.
\Pexpressive{} is also asymptotically faster than most arbitrary-choice printers, as it treats a document as a DAG rather than a tree.
Like Yelland's printer, for each layout under consideration, \Pexpressive{} keeps track of two quantities: cost and last line length. 
This is different from most printers in the family which keep track of three quantities: height, width, and last width.
The dimension reduction further makes \Pexpressive{} more efficient.
The concept of cost also allows \Pexpressive{} to decouple the page width limit and computation width limit, which allows graceful overflow handling.
\Pexpressive{}, unlike Yelland's printer, is parameterized by a cost factory, which supports a variety of optimality objectives without requiring a modification to the core printer.
This includes not only the linear optimality objectives that Yelland's printer supports, but also non-linear optimality objectives that can properly avoid overflows.

\subsection{Other Printers}

\citet{coutaz} introduced one of the earliest document abstractions for user interfaces.
The abstraction is very general: it can not only describe text layout, but also image and objects on computer screen.
Due to its minimality, it is much less expressive than other printers for textual printing.

\citet{hughes} brought pretty printing to the functional world. 
The work pioneers using combinators to construct a document for pretty printing, which is now a standard practice.
The printer targets a PPL that is neither the traditional nor arbitrary-choice PPL, but somewhere in-between. 
In particular, it only supports aligned concatenation and does not provide the arbitrary-choice operator in the public interface.
The work is more similar to the traditional printers in how it makes choices greedily, which minimizes neither overflow nor number of lines.
The combination of greedy choice making and aligned concatenation makes some documents print very poorly~\cite{bernardy}.
Furthermore, \citet{peyton} identified quadratic time complexity in the printer.

\citet{leijen} implemented Wadler's printer in Haskell and added support for aligned concatenation via the inclusion of \pp{align}, becoming the first printer that supports both aligned and unaligned concatenation.
However, similar to Hughes' printer, the printer can produce very poor output~\cite{bernardy}.

\section{An overview of \texorpdfstring{\Pexpressive}{Pi\ e}}\label{sec:tour}

\Pexpressive{} takes as inputs a document in \Lexpressive{}, a cost factory, and a computation width limit, and returns a textual layout.
This section provides an overview of \Pexpressive{} from the user's perspective---what form the inputs take, and how they interact to produce a layout.

\subsection{Documents in \texorpdfstring{\Lexpressive}{Sigma\ e}}\label{sec:tour:informal}

\begin{figure}
  \footnotesize
\begin{minipage}{.57\textwidth}
\begin{tabular}{l l@{\hspace{.5em}}l}
Document & $d \in \mathcal{D}_e\ \defn$ & 
\lstinline[language=pp]!text! $s \vbar$ 
\lstinline[language=pp]!nl! $\vbar$ 
$d$ \lstinline[language=pp]!<>! $d \vbar$  
\lstinline[language=pp]!nest! $ n\ d \vbar$ \\ 
& & 
\lstinline[language=pp]!align! $ d \vbar$  
\lstinline[language=pp]!flatten! $d \vbar$ 
$d$ \lstinline[language=pp]!<|>! $ d$  \\
\end{tabular}
\end{minipage}%
\begin{minipage}{.43\textwidth}
\begin{tabular}{l l@{\hspace{.5em}}l}
String without newline 	& $s, t, \ldots \in \mathsf{Str}$ &   \\
Natural number	& $n \in \mathbb{N}$ & \\
\end{tabular}
\end{minipage}
  \caption{Syntax for \Lexpressive{}}\label{fig:document-syntax}
\end{figure}

\begin{figure}
  \footnotesize
  \begin{subfigure}{.41\textwidth}
    \centering
    \begin{tabular}{c}
      \begin{lstlisting}[language=literal, mathescape]
     (!\tikzmark{gconcatenation-example1l}!)     (!\tikzmark{gconcatenation-example2r}!)   (!\tikzmark{gconcatenation-example3r}!)(&|&)
(!\tikzmark{gconcatenation-example4l}!)     (!\tikzmark{gconcatenation-example5l}!)  (!\tikzmark{gconcatenation-example6r}!)   (!\tikzmark{gconcatenation-example7r}!)(&|&)
(!\tikzmark{gconcatenation-example8l}!)     (!\tikzmark{gconcatenation-example9l}!)  (!\tikzmark{gconcatenation-example10r}!)(!\tikzmark{gconcatenation-example11l}!)      (!\tikzmark{gconcatenation-example12r}!)(&|&)
(!\tikzmark{gconcatenation-example13l}!)       (!\tikzmark{gconcatenation-example14l}!)    (!\tikzmark{gconcatenation-example15r}!)  (!\tikzmark{gconcatenation-example16r}!)(&|&)
(!\tikzmark{gconcatenation-example17l}!)           (!\tikzmark{gconcatenation-example18r}!)(&|&)
\end{lstlisting}
\begin{tikzpicture}[remember picture,overlay]
\fill[rounded corners=1pt, fill=blue, fill opacity=0.15]
([shift={(0pt,7pt)}]pic cs:gconcatenation-example1l)
--
([shift={(0pt,7pt)}]pic cs:gconcatenation-example2r)
--
([shift={(0pt,-2pt)}]pic cs:gconcatenation-example7r)
--
([shift={(0pt,-2pt)}]pic cs:gconcatenation-example6r)
--
([shift={(0pt,-2pt)}]pic cs:gconcatenation-example10r)
--
([shift={(0pt,-2pt)}]pic cs:gconcatenation-example8l)
--
([shift={(0pt,7pt)}]pic cs:gconcatenation-example4l)
--
([shift={(0pt,7pt)}]pic cs:gconcatenation-example5l)
--
([shift={(0pt,-2pt)}]pic cs:gconcatenation-example1l)
--
([shift={(0pt,7pt)}]pic cs:gconcatenation-example1l);
\fill[rounded corners=1pt, fill=yellow, fill opacity=0.2]
([shift={(0pt,7pt)}]pic cs:gconcatenation-example11l)
--
([shift={(0pt,7pt)}]pic cs:gconcatenation-example12r)
--
([shift={(0pt,-2pt)}]pic cs:gconcatenation-example16r)
--
([shift={(0pt,-2pt)}]pic cs:gconcatenation-example15r)
--
([shift={(0pt,-2pt)}]pic cs:gconcatenation-example18r)
--
([shift={(0pt,-2pt)}]pic cs:gconcatenation-example17l)
--
([shift={(0pt,7pt)}]pic cs:gconcatenation-example13l)
--
([shift={(0pt,7pt)}]pic cs:gconcatenation-example14l)
--
([shift={(0pt,-2pt)}]pic cs:gconcatenation-example11l)
--
([shift={(0pt,7pt)}]pic cs:gconcatenation-example11l);
\draw[thick, dashed]
([shift={(0pt,7pt)}]pic cs:gconcatenation-example1l)
--
([shift={(0pt,7pt)}]pic cs:gconcatenation-example3r)
--
([shift={(0pt,-2pt)}]pic cs:gconcatenation-example16r)
--
([shift={(0pt,-2pt)}]pic cs:gconcatenation-example15r)
--
([shift={(0pt,-2pt)}]pic cs:gconcatenation-example18r)
--
([shift={(0pt,-2pt)}]pic cs:gconcatenation-example17l)
--
([shift={(0pt,7pt)}]pic cs:gconcatenation-example4l)
--
([shift={(0pt,7pt)}]pic cs:gconcatenation-example5l)
--
([shift={(0pt,-2pt)}]pic cs:gconcatenation-example1l)
--
([shift={(0pt,7pt)}]pic cs:gconcatenation-example1l);
\draw
([shift={(0pt,7pt)}]pic cs:gconcatenation-example9l)node{$\cl{d}_a$};
\draw
([shift={(0pt,2pt)}]pic cs:gconcatenation-example14l)node[anchor=west]{$\cl{d}_b$};
\end{tikzpicture}
    \end{tabular}
    \caption{An unaligned concatenation of $\cl{d}_a$ and $\cl{d}_b$}\label{fig:semantics-illustration:concat}
  \end{subfigure}\hspace{2mm}
  \begin{subfigure}{.26\textwidth}
    \centering
    \begin{tabular}{c}
      \begin{lstlisting}[language=literal, mathescape]
(!\tikzmark{gnest-example1l}!)   (!\tikzmark{gnest-example2l}!)  (!\tikzmark{gnest-example3l}!)        (!\tikzmark{gnest-example4r}!)(&|&)
(!\tikzmark{gnest-example5l}!)   (!\tikzmark{gnest-example6l}!)  (!\tikzmark{gnest-example7l}!) (!\tikzmark{gnest-example8l}!)    (!\tikzmark{gnest-example9r}!)   (!\tikzmark{gnest-example10r}!)(&|&)
(!\tikzmark{gnest-example11l}!)   (!\tikzmark{gnest-example12l}!)       (!\tikzmark{gnest-example13r}!)(&|&)
\end{lstlisting}
\begin{tikzpicture}[remember picture,overlay]
\fill[rounded corners=1pt, fill=blue, fill opacity=0.15]
([shift={(0pt,7pt)}]pic cs:gnest-example3l)
--
([shift={(0pt,7pt)}]pic cs:gnest-example4r)
--
([shift={(0pt,-2pt)}]pic cs:gnest-example10r)
--
([shift={(0pt,-2pt)}]pic cs:gnest-example9r)
--
([shift={(0pt,-2pt)}]pic cs:gnest-example13r)
--
([shift={(0pt,-2pt)}]pic cs:gnest-example12l)
--
([shift={(0pt,7pt)}]pic cs:gnest-example6l)
--
([shift={(0pt,7pt)}]pic cs:gnest-example7l)
--
([shift={(0pt,-2pt)}]pic cs:gnest-example3l)
--
([shift={(0pt,7pt)}]pic cs:gnest-example3l);
\draw[<->]
([shift={(0pt,2pt)}]pic cs:gnest-example1l)
--
([shift={(0pt,2pt)}]pic cs:gnest-example2l)node[midway,above=-2pt]{\footnotesize $n$};
\draw[->]
([shift={(0pt,2pt)}]pic cs:gnest-example5l)
--
([shift={(0pt,2pt)}]pic cs:gnest-example6l);
\draw[->]
([shift={(0pt,2pt)}]pic cs:gnest-example11l)
--
([shift={(0pt,2pt)}]pic cs:gnest-example12l);
\draw[thick, dashed]
([shift={(0pt,7pt)}]pic cs:gnest-example3l)
--
([shift={(0pt,7pt)}]pic cs:gnest-example4r)
--
([shift={(0pt,-2pt)}]pic cs:gnest-example10r)
--
([shift={(0pt,-2pt)}]pic cs:gnest-example9r)
--
([shift={(0pt,-2pt)}]pic cs:gnest-example13r)
--
([shift={(0pt,-2pt)}]pic cs:gnest-example11l)
--
([shift={(0pt,7pt)}]pic cs:gnest-example5l)
--
([shift={(0pt,7pt)}]pic cs:gnest-example7l)
--
([shift={(0pt,-2pt)}]pic cs:gnest-example3l)
--
([shift={(0pt,7pt)}]pic cs:gnest-example3l);
\draw
([shift={(0pt,2pt)}]pic cs:gnest-example8l)node[anchor=west]{$\cl{d}$};
\end{tikzpicture}
    \end{tabular}
    \caption{A \pp{nest} of $\cl{d}$ to increase the indentation level by $n$}\label{fig:semantics-illustration:nest}
  \end{subfigure}\hspace{2mm}
  \begin{subfigure}{.29\textwidth}
    \centering
    \begin{tabular}{c}
      \begin{lstlisting}[language=literal, mathescape]
(!\tikzmark{galign-example1l}!)     (!\tikzmark{galign-example2l}!)          (!\tikzmark{galign-example3r}!)(&|&)
(!\tikzmark{galign-example4l}!)     (!\tikzmark{galign-example5l}!)  (!\tikzmark{galign-example6l}!)     (!\tikzmark{galign-example7r}!)   (!\tikzmark{galign-example8r}!)(&|&)
(!\tikzmark{galign-example9l}!)     (!\tikzmark{galign-example10l}!)       (!\tikzmark{galign-example11r}!)(&|&)
\end{lstlisting}
\begin{tikzpicture}[remember picture,overlay]
\fill[rounded corners=1pt, fill=blue, fill opacity=0.15]
([shift={(0pt,7pt)}]pic cs:galign-example2l)
--
([shift={(0pt,7pt)}]pic cs:galign-example3r)
--
([shift={(0pt,-2pt)}]pic cs:galign-example8r)
--
([shift={(0pt,-2pt)}]pic cs:galign-example7r)
--
([shift={(0pt,-2pt)}]pic cs:galign-example11r)
--
([shift={(0pt,-2pt)}]pic cs:galign-example10l)
--
([shift={(0pt,-2pt)}]pic cs:galign-example2l)
--
([shift={(0pt,7pt)}]pic cs:galign-example2l);
\draw[->]
([shift={(0pt,2pt)}]pic cs:galign-example4l)
--
([shift={(0pt,2pt)}]pic cs:galign-example5l);
\draw[->]
([shift={(0pt,2pt)}]pic cs:galign-example9l)
--
([shift={(0pt,2pt)}]pic cs:galign-example10l);
\draw[thick, dashed]
([shift={(0pt,7pt)}]pic cs:galign-example2l)
--
([shift={(0pt,7pt)}]pic cs:galign-example3r)
--
([shift={(0pt,-2pt)}]pic cs:galign-example8r)
--
([shift={(0pt,-2pt)}]pic cs:galign-example7r)
--
([shift={(0pt,-2pt)}]pic cs:galign-example11r)
--
([shift={(0pt,-2pt)}]pic cs:galign-example9l)
--
([shift={(0pt,7pt)}]pic cs:galign-example4l)
--
([shift={(0pt,7pt)}]pic cs:galign-example5l)
--
([shift={(0pt,-2pt)}]pic cs:galign-example2l)
--
([shift={(0pt,7pt)}]pic cs:galign-example2l);
\draw
([shift={(0pt,2pt)}]pic cs:galign-example6l)node[anchor=west]{$\cl{d}$};
\draw[<->]
([shift={(0pt,2pt)}]pic cs:galign-example1l)
--
([shift={(0pt,2pt)}]pic cs:galign-example2l)node[midway,above=-2pt]{\footnotesize $c$};
\end{tikzpicture}
    \end{tabular}
    \caption{An \pp{align} of $\cl{d}$ when rendered at the column position $c$}\label{fig:semantics-illustration:align}
  \end{subfigure}
  \caption{Illustrations of constructs in \Lexpressive{}.
  The area with dashed borders is the resulting layout.}\label{fig:semantics-illustration}
\end{figure}

Like other printers, \Pexpressive{} allows users to construct a document to encode a structured data along with formatting choices.
The document can be \emph{evaluated} to a set of layouts, and
\Pexpressive{} will pick an optimal layout from this set as the output.

The document is written in the \Lexpressive{} syntax, shown in \Cref{fig:document-syntax}.
Each construct is from either the traditional or the arbitrary-choice PPLs, except for the \pp{flatten} construct (which is used internally in \citet{wadler}'s printer) and the \pp{align} construct (which is from \citet{leijen}'s printer).

For now, we ignore the (arbitrary-) choice operator \pp{<|>}.
A document without the choice operator is called a \emph{choiceless document}, denoted by $\cl{d} \in \CLDocT$.
A choiceless document can be \emph{rendered} at a \emph{column position} $c$ and an \emph{indentation level} $i$ (both default to 0) to produce a single layout.
The informal semantics of choiceless document are as follows:

\vspace{2mm}
\begin{tabular}{p{\dimexpr .13\textwidth-2\tabcolsep}p{\dimexpr .84\textwidth-2\tabcolsep}}
\lstinline[language=pp, mathescape]|text $s$| 
& renders to a layout with a single line $s$. \\
\end{tabular}

\begin{tabular}{p{\dimexpr .13\textwidth-2\tabcolsep}p{\dimexpr .84\textwidth-2\tabcolsep}}
\lstinline[language=pp]!nl! & normally renders to a layout with two lines. 
The first line is empty, and the second line consists of $i$ spaces.
  \lstinline[language=pp]!nl! interacts with \emph{flattening}, which reduces it to just a single space.\\
\end{tabular}

\begin{tabular}{p{\dimexpr .13\textwidth-2\tabcolsep}p{\dimexpr .84\textwidth-2\tabcolsep}}
\lstinline[language=pp, mathescape]!$\cl{d}_a$ <> $\cl{d}_b$! 
& renders to a layout that concatenates the layout of $\cl{d}_a$ and the layout of $\cl{d}_b$.
This is the \emph{unaligned} concatenation from the traditional PPL, illustrated in \Cref{fig:semantics-illustration:concat}.\\
\end{tabular}

\begin{tabular}{p{\dimexpr .13\textwidth-2\tabcolsep}p{\dimexpr .84\textwidth-2\tabcolsep}}
  \lstinline[language=pp, mathescape]!nest $n$ $\cl{d}$! & renders to a layout like $\cl{d}$, but with the indentation level $i$ relatively \emph{increased} by $n$. 
  \Cref{fig:semantics-illustration:nest} illustrates this.\\
\end{tabular}

\begin{tabular}{p{\dimexpr .13\textwidth-2\tabcolsep}p{\dimexpr .84\textwidth-2\tabcolsep}}
\lstinline[language=pp, mathescape]!align $\cl{d}$! 
  & renders to a layout like $\cl{d}$, but with alignment: the indentation level $i$ is \emph{set} (not relatively increased) to the column position $c$.
\Cref{fig:semantics-illustration:align} illustrates this.\\
\end{tabular}

\begin{tabular}{p{\dimexpr .13\textwidth-2\tabcolsep}p{\dimexpr .84\textwidth-2\tabcolsep}}
\lstinline[language=pp, mathescape]!flatten $\cl{d}$! 
& renders to a layout like $\cl{d}$, but with all newlines and indentation spaces due to \lstinline[language=pp]!nl!s flattened to single spaces.\\
\end{tabular}
\vspace{2mm}

\begin{example}\label{ex:choiceless-render}
  When the following choiceless document is rendered at column position $3$ and indentation level $0$, it produces the second layout in \Cref{fig:cost-factory-illustration}:
  \begin{lstlisting}[language=pp]
text "= func(" <> nest 2 (nl <> text "pretty," <> nl <> text "print") <> nl <> text ")"
  \end{lstlisting} 
\end{example}

While \Cref{fig:semantics-illustration} provides a rough illustration that should be helpful to understand the semantics of choiceless document, it could be misleading, as shown in the next example.

\begin{example}
The document \pp{text "a" <> (nest 42 (align (text "b" <> nl <> text "c")))} is rendered at the column position and indentation level 0 to a layout with two lines: \pp{"ab"} and \pp{" c"}.
The nesting doesn't visibly increase the indentation level by 42. 
To see why, note that \pp{nest 42 ...} is rendered at the column position 1 and indentation level 0.
Subsequently, \pp{align ...} is rendered at the column position 1 and indentation level 42.
Then, \pp{text "b" <> nl <> text "c"} is rendered at the column position 1 and indentation level 1.
That is, the alignment on the inner document overrides the indentation level.
This example shows the importance of the indentation level, and why it must be specifically tracked.
\end{example}

This concludes our informal description of how a choiceless document renders to a layout. 
General documents, by contrast, can contain the (arbitrary-) choice operator \pp{<|>}, which provides a choice among the layouts from two sub-documents.
Thus, unlike choiceless documents, which render to a single layout, general documents will \emph{evaluate} to a non-empty, finite set of layouts.
Intuitively, this is done by first \emph{widening} a document into a set of choiceless documents, then rendering each choiceless document in the set, producing a set of layouts.

\begin{example}
  The document \pp{(text "a" <|> text "b") <> (text "c" <|> text "d")} widens to four choiceless documents: \pp{text "a" <> text "c"}, \pp{text "a" <> text "d"}, \pp{text "b" <> text "c"}, and \pp{text "b" <> text "d"}.
  Thus, the document evaluates (with column position and indentation level 0) to a set of four layouts: \pp{"ac"}, \pp{"ad"}, \pp{"bc"}, and \pp{"bd"}.
\end{example}

\subsection{Cost Factory}\label{sec:tour:cost-factory}

\begin{figure}
  
\footnotesize
\begin{tabular}{r @{ : } l p{\dimexpr .75\textwidth-2\tabcolsep} }
\multicolumn{1}{r}{} & Cost type $\tau$  \\
\cfle & $\tau \to \tau \to \B$ & \cfle{} must be a total ordering (transitive, antisymmetric, and total) \\
\cfcombine & $\tau \to \tau \to \tau$ & 
$\forall \mathcal{C}_1, \mathcal{C}_2, \mathcal{C}_3, \mathcal{C}_4 \in \tau .\ [\mathcal{C}_1 \cfle \mathcal{C}_2 \to \mathcal{C}_3 \cfle \mathcal{C}_4 \to \mathcal{C}_1 \cfcombine \mathcal{C}_3 \cfle \mathcal{C}_2 \cfcombine \mathcal{C}_4]$ \\
\cftext & $\N \to \N \to \tau$ & 
$\forall c, c', l \in \N .\ [c \le c' \to \cftext(c, l) \cfle \cftext(c', l)]$ \\
$\cfnl$ & $\tau$ & $\forall c, l_1, l_2 \in \N .\ \cftext(c, l_1 + l_2) = \cftext(c, l_1) \cfcombine \cftext(c + l_1, l_2)$\\
\multicolumn{2}{c}{} & $\cfcombine$ must be associative, with the identity that is $\cftext(0, 0)$\\
\multicolumn{2}{c}{} & $\forall c \in \N .\ \cftext(c, 0) = \cftext(0, 0)$\\
\end{tabular}

  \caption{The cost factory interface.
  Users need to supply the cost type $\tau$ and implement the operations satisfying the contracts indicated in the interface.}\label{fig:cost-factory}
\end{figure}

\begin{figure}
  \parbox{\figrasterwd}{
  \parbox{.48\figrasterwd}{%
    \begin{lstlisting}[numbers=left, numbersep=.75em, language=literal, xleftmargin=1.5em]
(!\tikzmark{gfactory-example-horz1l}!)   (!\tikzmark{gfactory-example-horz2l}!)= f(!\tikzmark{gfactory-example-horz3l}!)unc( pre(!\tikzmark{gfactory-example-horz4l}!)tty, print )(&|&)
\end{lstlisting}
\begin{tikzpicture}[remember picture,overlay]
\draw[red, ultra thick, densely dashed]
([shift={(0pt,7pt)}]pic cs:gfactory-example-horz3l)
--
([shift={(0pt,-2pt)}]pic cs:gfactory-example-horz3l);
\draw[<->]
([shift={(0pt,2pt)}]pic cs:gfactory-example-horz1l)
--
([shift={(0pt,2pt)}]pic cs:gfactory-example-horz2l)node[midway,above=-2pt]{\footnotesize 3};
\draw[teal, ultra thick, densely dashed]
([shift={(0pt,7pt)}]pic cs:gfactory-example-horz4l)
--
([shift={(0pt,-2pt)}]pic cs:gfactory-example-horz4l);
\end{tikzpicture}
    \vspace{-4mm}
  }
  \hskip1em
  \parbox{.48\figrasterwd}{%
      \begin{lstlisting}[numbers=left, numbersep=.75em, language=literal, xleftmargin=1.5em]
(!\tikzmark{gfactory-example-vert1l}!)   (!\tikzmark{gfactory-example-vert2l}!)= f(!\tikzmark{gfactory-example-vert3l}!)unc(    (!\tikzmark{gfactory-example-vert4l}!) (&|&)
  pretty,(&|&)
  print(&|&)
)     (!\tikzmark{gfactory-example-vert5l}!)        (!\tikzmark{gfactory-example-vert6l}!) (&|&)
\end{lstlisting}
\begin{tikzpicture}[remember picture,overlay]
\draw[red, ultra thick, densely dashed]
([shift={(0pt,7pt)}]pic cs:gfactory-example-vert3l)
--
([shift={(0pt,-2pt)}]pic cs:gfactory-example-vert5l);
\draw[<->]
([shift={(0pt,2pt)}]pic cs:gfactory-example-vert1l)
--
([shift={(0pt,2pt)}]pic cs:gfactory-example-vert2l)node[midway,above=-2pt]{\footnotesize 3};
\draw[teal, ultra thick, densely dashed]
([shift={(0pt,7pt)}]pic cs:gfactory-example-vert4l)
--
([shift={(0pt,-2pt)}]pic cs:gfactory-example-vert6l);
\end{tikzpicture}
      \vspace{-4mm}
  }
  }
  \scriptsize
  \caption{Two example layouts to illustrate how a cost factory computes their costs. Both layouts are rendered at column position 3. The dotted lines shows the width limit of 6 and 14.}
  \label{fig:cost-factory-illustration}
\end{figure}

To pick an optimal layout from the set of layouts that a document evaluates to, \Pexpressive{} needs to be able to compute a cost for each layout, and to compare these costs to find a layout with minimal cost.
To accommodate a wide range of optimality objectives, we allow the user to specify a cost type $\tau$ and implement operations on the cost type:

\begin{itemize}
  \item a procedure $\cftext(c, l)$ that computes the cost of text starting at column $c$ of length $l$
  \item a constant $\cfnl$ that gives the cost of a newline\footnote{In our Lean formalization and actual implementation, $\cfnl$ is a procedure. See \Cref{sec:impl} for details.}
  \item a procedure $\cfcombine$ that adds two costs together
  \item a procedure $\cfle$ that compares two costs
\end{itemize}

We call this set of parameters a \emph{cost factory}.
These parameters cannot be arbitrary, however.
For example, the cost of \pp{"hello world"} placed at column 10 should be the same as the cost of \pp{"hello "} placed at column 10 combined with the cost of \pp{"world"} placed at column 16.\footnote{In other words, the cost of a long text should be able to be broken down into the costs of its characters.}
Thus, a \emph{valid} cost factory also needs to additionally satisfy the contracts listed in \Cref{fig:cost-factory}.\footnote{For mathematical readers, a (valid) cost factory forms a totally ordered monoid with translational invariance.}
The first three contracts allow \Pexpressive{} to efficiently prune away suboptimal costs during incremental cost computation (\Cref{sec:printer}), and the last three contracts ensure that the concept of the cost for a layout is well-defined.

With a cost factory, we can inductively compute the cost of a layout with lines $l_{1}, l_{2}, \ldots, l_{n}$ rendered at column position $c$:
\begin{align*}
\textsc{Cost}([l_{1}], c) &= \cftext(c, |l_{1}|)\\
\textsc{Cost}([l_{1}, l_{2}, \ldots, l_{n-1}, l_{n}], c) &=
 \textsc{Cost}([l_{1}, l_{2}, \ldots, l_{n-1}], c) \cfcombine \cfnl \cfcombine \cftext(0, |l_{n}|)
\end{align*}

\Pexpressive{} can then use $\cfle$ to find an optimal layout.
The following example shows a concrete cost factory and how it can be used to pick an optimal layout among the layouts in \Cref{fig:cost-factory-illustration}.

\begin{example}\label{ex:cost-factory-linear}
  Consider an optimality objective that minimizes the sum of \emph{overflows} (the number of characters that exceed a given \emph{page width limit} $w$ in each line), and \emph{then} minimizes the \emph{height} (the total number of newline characters, or equivalently, the number of lines minus one).
  This objective is thus able to avoid the excessive overflow problem in Bernardy's printer described in \Cref{sec:related}.

  More concretely, the cost of a layout is a pair of the overflow sum and the height, where lexicographic order determines which cost is less.
  With $w = 6$, the first layout in \Cref{fig:cost-factory-illustration} has the cost $(20, 0)$, whereas the second layout has the cost $(4 + 3 + 1 + 0, 3) = (8, 3)$.
  Thus, the second layout is the optimal layout that \Pexpressive{} should pick.

  We implement this optimality objective with the following cost factory $\mathcal{F}$.
  \[
    \begin{aligned}
    \tau &= \N \times \N &
    \cfle &=\ \le_{\mathsf{lex}} &
    (o_a, h_a) \cfcombine (o_b, h_b) &= (o_a + o_b, h_a + h_b) \\
  \end{aligned}
  \] \[
    \begin{aligned}
    \cftext(c, l) &=
    (\max(c + l - \max(w, c), 0), 0) &
    \cfnl &= (0, 1)
  \end{aligned}
  \]

  According to $\mathcal{F}$, the first layout has cost  $\cftext(3, 26) = (20, 0)$, while the second layout has the cost $\cftext(3, 7) \cfcombine \cfnl \cfcombine \cftext(0, 9) \cfcombine \cfnl \cfcombine \cftext(0, 7) \cfcombine \cfnl \cfcombine \cftext(0, 1) = (8, 3)$, as expected.
\end{example}

The cost factory interface is versatile.
The above example shows that \Pexpressive{} does not need to take a page width limit as an input, because the concept of page width limit can already be defined by users via \cftext.
It is also possible, for example,
to implement \emph{soft} width limits, or to 
compute a linear combination of height and overflow in the style of \citet{yelland}.
The rest of this section provides a couple more examples of other valid and invalid cost factories.

\begin{example}\label{ex:cost-factory-quadratic}
  The following cost factory targets an optimality objective that minimizes the sum of \emph{squared} overflows over the page width limit $w$, and then the height.
  This optimality objective is an improvement over the one in \Cref{ex:cost-factory-linear} by discouraging overly large overflows.
  With $w = 6$, the first layout in \Cref{fig:cost-factory-illustration} has the cost $(20^2, 0)$ whereas the second layout has the cost $(4^{2} + 3^{2} + 1^{2} + 0^2, 3)$
  The text cost formula is derived from the identity $(a + b)^2 - a^2 = b(2a+b)$ where in each text placement, $a$ is the starting position count past $w$ and $b$ is the overflow length.
  This is (essentially) the default cost factory that our implementation, \prettiester{}, employs.
  \[ 
    \begin{aligned}
    \tau &= \N \times \N &
    \cfle &=\ \le_{\mathsf{lex}} &
    (o_a, h_a) \cfcombine (o_b, h_b) &= (o_a + o_b, h_a + h_b)  &
    \cfnl &= (0, 1) \\
  \end{aligned}
  \] \[
    \begin{aligned}
   \cftext(c, l) &= \begin{cases}
       (b(2a + b), 0) & \text{if } c + l > w \\
       (0, 0) & \text{otherwise}
   \end{cases} &
   \text{where} \quad & \begin{aligned}
       a &= \max(w, c) - w \\
       b &= c + l - \max(w, c)
   \end{aligned}
  \end{aligned}
  \]
\end{example}

\begin{example}\label{ex:cost-factory-max}
  The following cost factory targets an optimality objective that minimizes the maximum overflow over the page width limit $w$.
With $w = 6$, the first layout in \Cref{fig:cost-factory-illustration} has the cost $20$ whereas the second layout has the cost $\max(4, 3, 1, 0) = 4$.
  \[
    \begin{aligned}
    \tau &= \N &
    \cfle &=\ \le &
    m_a \cfcombine m_b &= \max(m_a, m_b)  &
    \cfnl &= 0 \\
  \end{aligned}
  \] \[
   \cftext(c, l) = \begin{cases}
       0 & \text{if } l = 0 \\
       \max(0, c + l - w) & \text{otherwise}
   \end{cases}
   \]
\end{example}

The above cost factories are all valid.
This is proven with automated theorem proving via Rosette 4~\cite{rosette4, rosette:pldi} and Z3~\cite{z3}.

\begin{theorem}
  The cost factories in \Cref{ex:cost-factory-linear}, \Cref{ex:cost-factory-quadratic}, and \Cref{ex:cost-factory-max} are valid.
\end{theorem}

\begin{example}\label{ex:cost-factory-max-lex}
  The following \emph{invalid} cost factory intends to target an optimality objective that minimizes the maximum overflow over the page width limit $w$, and then the height.
  However, the second contract is violated, because $(0, 1) \cfcombine (2, 0) \cfle (1, 0) \cfcombine (2, 0)$ does not hold.
  \[
    \begin{aligned}
    \tau &= \N \times \N &
    \cfle &=\ \le_{\mathsf{lex}} &
    (m_a, h_{a}) \cfcombine (m_b, h_{b}) &= (\max(m_a, m_b), h_{a} + h_{b})  &
    \cfnl &= (0, 1) \\
  \end{aligned}
  \] \[
   \cftext(c, l) = \begin{cases}
       (0, 0) & \text{if } l = 0 \\
       (\max(0, c + l - w), 0) & \text{otherwise}
   \end{cases}
   \]
\end{example}

\subsection{\texorpdfstring{\cfw}{W}, the Computation Width Limit}\label{sec:tour:limit}

The last input to \Pexpressive{} is \cfw{}, the computation width limit.
When printing a document $d$, \Pexpressive{} only provides the optimality guarantee among layouts evaluated from $d$ whose column position or indentation level during the printing process does not exceed $\cfw$.
For each choiceless document widened from $d$, when its rendering causes a column position or indentation level to exceed the computation width limit, the rendering is \emph{tainted}.
  For example, if a document evaluates to two layouts in \Cref{fig:cost-factory-illustration}, with $\cfw = 14$, the rendering to the first layout would be tainted, while the rendering to the second layout would not (assuming the indentation level during the rendering doesn't exceed the limit).
  Layouts from tainted rendering can usually be discarded right away, except when every possible rendering is tainted.
  In such case, \Pexpressive{} keeps one layout so that it can still output a layout, but provides no guarantee that the layout will be optimal.
  The tainting system allows us to bound the computation so that the algorithm is efficient.

\section{The semantics of \texorpdfstring{\Lexpressive}{Sigma\ e}}\label{sec:semantics}

This section formally presents \Lexpressive{}, an expressive PPL.
We begin this section by describing \emph{layouts}, which are the textual outputs. 
Then, we formally describe the semantics of \Lexpressive{}, which is determined by the evaluation of a document in \Lexpressive{} to a set of layouts.

\subsection{Layouts}\label{sec:layout}

A \emph{layout} $l \in \LayoutT$ is a textual output. 
We represent a layout as a non-empty, finite list of lines (implicitly joined by newlines), where each line is a string without the newline character.\footnote{The representation in our Lean formalization is more elaborated, making indentation level explicit by incorporating the information into a layout. We present a simplified version here for the sake of simplicity.
See \Cref{sec:impl} for details.}
This allows us to easily reason about the number of lines and the length of each line.
The first line of a layout can be put at an arbitrary column position (depending on which column position it is rendered at), but subsequent lines must be put at the column position $0$.

\begin{example}
  The second layout in \Cref{fig:cost-factory-illustration} is \lstinline[language=pp,mathescape]{$[$"func("$,\ $"  pretty,"$,\ $"  print"$,\ $")"$]$}, which is rendered at the column position 3.
\end{example}

\subsection{The Formal Semantics of \texorpdfstring{\Lexpressive}{Sigma\ e}}\label{sec:semantics:formal}

\begin{figure}
  \input{fig/document-semantics}
  \caption{Semantics for \Lexpressive{}. 
  ``$\epsilon$'' is the empty string. 
  ``$s \times i$'' is the notation for replicating the string $s$ for $i$ times. 
  ``$s \stringconcat t$'' is a string concatenation of $s$ and $t$.
  Lastly, ``$s_1, \ldots, s_n$'' and ``$s_1, \ldots^+, s_n$''  indicate $n$ lines, where $n \ge 0$ and $n \ge 1$ respectively.}\label{fig:document-semantics}
\end{figure}

Our approach to evaluate a document in \Lexpressive{} to a set of layouts is to first \emph{widen} a document into a set of choiceless documents, then render each choiceless document in the set, producing a set of layouts.

The formal semantics of \Lexpressive{} consists of two relations, shown in \Cref{fig:document-semantics}.
The judgment \renderL{\cl{d}}{c}{i}{f}{l} states that the choiceless document $\cl{d} \in \CLDocT$ placed at column position $c \in \N$ with indentation level $i \in \N$ and flattening mode $f \in \B$, will render to the layout $l \in \LayoutT$.
Unlike the informal semantics in \Cref{sec:tour:informal}, we make the flattening mode $f$, which indicates whether newlines should be replaced with spaces, explicit.
Its value can be either on ($\top$) or off ($\bot$).
Another judgment \widen{d}{\overline{D}} states that a document $d \in \mathcal{D}_e$ is widened to a finite, non-empty set of choiceless documents $\overline{D} \in 2^{\CLDocT}$.
We sometimes call a combination of $c$ and $i$ (and possibly $f$) a \emph{printing context}.
Now, we elaborate some interesting rules in the figure.

\paragraph{Rendering Text} 
The \textsc{Text} rule states that the rendering of a text placement \lstinline[language=pp, mathescape]{text $s$} contains a layout with a single line of the text $s$.
The printing context is completely ignored.

\paragraph{Rendering Newlines}
When the flattening mode is off, the \textsc{LineNoFlatten} rule states that the rendering of a \pp{nl} results in a layout with two lines.
The first line is empty, while the second line is indented by $i$ spaces.
On the other hand, when the flattening mode is on, 
the \textsc{LineFlatten} rule states that the rendering of the newline results in a layout with a single line of a single space.

\paragraph{Rendering Unaligned Concatenation}
In the rendering of \lstinline[language=pp, mathescape]{$\cl{d}_a$ <> $\cl{d}_b$}, we recursively render $\cl{d}_a$ and $\cl{d}_b$, but the rendering of $\cl{d}_b$ is dependent on the rendering of $\cl{d}_a$.
Let $l_a$ be the rendering result of $\cl{d}_a$.
The \textsc{ConcatOne} rule handles the case where $l_a$ has a single line, and the \textsc{ConcatMult} rule handles the case where $l_a$ has multiple lines. 
\begin{itemize}
    \item{If $l_a$ has only a single line $s$, the column position of $\cl{d}_b$'s rendering needs to be after the string $s$ is placed, i.e. at $c + |s|$.
    In such case, let $l_b$ be the rendering result of $\cl{d}_b$. 
    The first line of the resulting layout is the concatenation of $s$ and the first line of $l_b$. 
    The rest of the lines are from the rest of $l_b$.}
    \item{On the other hand, if $l_a$ has multiple lines, the column position of $d_b$'s rendering is simply the column position after the last line is placed.
    In such case, let $l_b$ be the rendering result of $\cl{d}_b$, the resulting layout contains all but the last line of $l_a$, a concatenation of the last line of $l_a$ and the first line of $l_b$, and the rest of $l_b$.}
\end{itemize}

\paragraph{Widening Choices}
The \textsc{UnionWiden} rule states that the widening of \lstinline[language=pp, mathescape]{$d_a$ <|> $d_b$} is the union of widen $d_a$ and widen $d_b$ 

\vspace{1em}

Both $\Downarrow_{\mathcal{R}}$ and $\Downarrow_{\mathcal{W}}$ are deterministic and total.
Thus, we can define $\render_e(d) = \set{l : \renderL{\cl{d}}{0}{0}{\bot}{l}, \cl{d} \in \cl{D}, \widen{d}{\cl{D}}}$ as the evaluation function for \Lexpressive{}, which consumes a document, widens it, and produces a set of layouts.

\section{A framework to reason about expressiveness}\label{sec:framework}

In previous sections, we informally made claims about expressiveness of PPLs. 
This section presents a framework to formally reason about it, based on two notions: \emph{functional completeness} and \emph{definability}.
We first define the semantics of the traditional and arbitrary-choice PPLs.
Then, we define our framework, and show that \Lexpressive{} is strictly more expressive than both the traditional and arbitrary-choice PPLs while being minimal.

In particular, \Cref{thm:expressive} states that every construct in the traditional and arbitrary-choice PPLs is definable in \Lexpressive{}.
However, \Cref{thm:inexpressibility-old-langs} states that some of these constructs are not definable in the traditional and arbitrary-choice PPLs.
Finally, \Cref{thm:minimal} shows that \Lexpressive{} is minimal.
Proof sketches of theorems in this section are provided in \appendixref{sec:proofs}.

\subsection{The Extended Semantics}

To reason about the traditional and arbitrary-choice PPLs, we need to precisely define their semantics.
To do so, we construct a PPL $\Sigma_{\mathsf{all}}$ that contains all constructs from \Lexpressive{} and the traditional and arbitrary-choice PPLs by extending \Cref{fig:document-semantics} with \Cref{fig:semantics-extension} (along with the straightforward widening rules).
Note that we follow \citet{wadler}'s approach by treating $\text{\pp{group} }d$ as a syntactic sugar for $d \text{ \pp{<|> flatten} } d$.
As \pp{<|>} and \pp{flatten} are already in $\Sigma_{\mathsf{all}}$, we do not need to
adjust anything further.

The extended semantics are still deterministic and total. 
The semantics of the traditional and arbitrary-choice PPLs is then the restricted semantics of $\Sigma_{\mathsf{all}}$ that only allows their constructs.%
\footnote{It is worth noting that there are many ways to specify rules to be consistent with the intended semantics of the arbitrary-choice PPL.
For instance, an invariant in the the arbitrary-choice PPL is that $c = i$ throughout the rendering process.
As a result, we could substitute the $\textsc{VertConcatNoFlatten}$ rule with its variant that changes the premise $\renderL{\cl{d}_b}{i}{i}{\bot}{[t_1, \ldots^+, t_m]}$ to $\renderL{\cl{d}_b}{c}{c}{\bot}{[t_1, \ldots^+, t_m]}$, without affecting the semantics of the arbitrary-choice PPL.
However, this change could affect the semantics of $\Sigma_{\mathsf{all}}$ and subsequent theorems in this section.
We pick $\textsc{VertConcatNoFlatten}$ over the variant because it seemingly integrates better with other constructs in $\Sigma_{\mathsf{all}}$.}
Throughout this section, we assume that any PPL is similarly a sublanguage of $\Sigma_{\mathsf{all}}$, whose semantics is well-defined and consistent with $\Sigma_{\mathsf{all}}$.

\begin{figure}
    \input{fig/semantics-extension}
    \caption{The semantics extension.}\label{fig:semantics-extension}
\end{figure}

\subsection{Functional Completeness}

In \Cref{sec:intro}, we claimed that the traditional PPL cannot express the two layouts in \Cref{fig:append:arbitrary},
as one layout has an extra pair of parentheses.
The question that we may want to ask in general then is, given a PPL $\Sigma$ and a non-empty set of layouts $L$, is it possible to construct a document in $\Sigma$ that evaluates to $L$? 
This motivates us to define the notion of functional completeness for PPLs.

\begin{definition}
    A PPL $\Sigma$ with an evaluation function $\render(\cdot)$ is \emph{functionally complete} if for any non-empty set of layouts $L$, there exists a document $d$ in $\Sigma$ such that $\render(d) = L$.
\end{definition}

With this definition, we can formally reason about some PPLs that we have previously seen.

\begin{restatable}{lemma}{arbitrarylexpressivefunctional}\label{lem:arbitrary-lexpressive-functional}
    The arbitrary-choice PPL and \Lexpressive{} are functionally complete.
\end{restatable}

\begin{restatable}{lemma}{traditionalfunctional}\label{lem:traditional-functional}
    The traditional PPL is \emph{not} functionally complete.
\end{restatable}

\begin{restatable}{lemma}{sublexpressivenotfunctional}\label{lem:sub-lexpressive-not-functional}
    For each construct $\mathbf{F}$ in $\set{\text{\pp{text}}, \text{\pp{<>}}, \text{\pp{nl}}, \text{\pp{<|>}}}$, \Lexpressive{} without $\mathbf{F}$ is not functionally complete.
\end{restatable}

If we limit the notion of expressiveness to only functional completeness, then all functionally complete PPLs would be equally expressive.
However, intuitively this is clearly not the case. 
The proof of \Cref{lem:arbitrary-lexpressive-functional} in \appendixref{sec:proofs} shows that it suffices for a PPL to only have \pp{text}, \pp{<$>}, and \pp{<|>} for functional completeness,
yet such a PPL would not be pleasant to use compared to \Lexpressive{}, because of the lack of features to, e.g., adjust indentation level.
In a sense, functional completeness for PPLs is similar to Turing completeness for programming languages, which similarly does not fully capture expressiveness for programming languages.
The next subsection presents a more fine-grained notion of expressiveness, based on the ability to define features.

\subsection{Definability}\label{sec:framework:definability}

The proof of \Cref{lem:arbitrary-lexpressive-functional} shows that while \Lexpressive{} doesn't have \pp{<$>}, we can simply expand $d_a\text{ \pp{<$>} }d_b$ to $d_a\text{ \pp{<> nl <>} }d_b$, which are in \Lexpressive{}, to perform the same functionality. 
In other words, the construct \pp{<$>} is already \emph{definable} by \pp{<>} and \pp{nl}.
Thus, adding \pp{<$>} to \Lexpressive{} doesn't increase its expressiveness. 
By contrast, \pp{<>} is not definable by any combination of features in the arbitrary-choice PPL.
To achieve the functionality of \pp{<>}, it would require a non-local restructuring of the document, making it difficult to construct the document in natural way.
In this sense, the inability to define a construct in a PPL means that adding the construct to the PPL increases its expressiveness.

More concretely, consider the document in the arbitrary-choice PPL shown in \Cref{fig:append:arbitrary}.
The document is awkwardly constructed.
The \lstinline[language=literal]{return} keyword must be distributed to combine with a first line of the returned expression, due to the undefinability of unaligned concatenation.
This creates a disconnection between the document structure and the underlying AST structure, making it more tedious and error-prone to construct documents.
By contrast, the following document is a rewrite of \Cref{fig:append:arbitrary} to utilize the full expressiveness of \Lexpressive{} in a natural way.
The sub-document colored blue fully corresponds to the ``returned expression,'' allowing users to recursively construct documents naturally.
\vspace{1mm}
\begin{lstlisting}[language=pp]
text "function append(first,second,third){" <> nest 4 ((&|&)
  let f = text "first +" in let s = text "second +" in let t = text "third" in(&|&)
  nl <> text "return " <>(&|&)
  (!\tikzmark{glexpressive-code1l}!)((text "(" <> (nest 4 (nl <> (f <> nl <> s <> nl (!\tikzmark{glexpressive-code2r}!)<> t))) <> nl <> text ")") <|>(!\tikzmark{glexpressive-code3r}!)(&|&)
  (!\tikzmark{glexpressive-code4l}!)  let sp = text " " in (f <> sp <> s <> sp <> t))(!\tikzmark{glexpressive-code5r}!)(&|&)
) <> nl <> text "}"(&|&)
\end{lstlisting}
\begin{tikzpicture}[remember picture,overlay]
\fill[rounded corners=1pt, fill=blue, fill opacity=0.15]
([shift={(0pt,7pt)}]pic cs:glexpressive-code3r)
--
([shift={(0pt,-2pt)}]pic cs:glexpressive-code3r)
--
([shift={(0pt,-2pt)}]pic cs:glexpressive-code2r)
--
([shift={(0pt,-2pt)}]pic cs:glexpressive-code5r)
--
([shift={(0pt,-2pt)}]pic cs:glexpressive-code4l)
--
([shift={(0pt,7pt)}]pic cs:glexpressive-code1l)
--
([shift={(0pt,7pt)}]pic cs:glexpressive-code3r);
\end{tikzpicture}

\vspace{-4mm}

The notion of definability (also known as expressibility) for programming languages was first developed by \citet{felleisen}, and we adapt it for PPLs through a series of definitions as follows:

\begin{definition}
    A PPL $\Sigma$ consists of:
    \begin{itemize}
    \item a set of (possibly infinitely many) function symbols $\Sigma = \set{\mathbf{F}, \ldots}$.
    The function symbols are referred to as \emph{construct}s. 
    Each may have different arity, argument sorts, and resulting sort.
    \item a non-empty set of documents $\mathcal{D}$ generated from $\Sigma$, where a document is a term of sort $\mathsf{Doc}$.
    \item an evaluation function $\render : \mathcal{D} \to 2^{\LayoutT}$.
    \end{itemize}
\end{definition}

\begin{example}
    \Lexpressive{} contains 
    \pp{nest}, 
    which is a construct with arity 2 of resulting sort \DocS.
    The first argument to \pp{nest} has sort $\mathbb{N}$ and the second argument has sort \DocS.
    \Lexpressive{} also contains all natural numbers and strings with no newline, which are constructs with arity 0 of resulting sort $\N$ and \StrT{} respectively.
    The evaluation function for \Lexpressive{} is $\render_e$ from \Cref{sec:semantics:formal}.
\end{example}

Henceforth, unless indicated otherwise, $\mathcal{D}_X$ and $\render_X$ are the set of documents and the evaluation function for the PPL $\Sigma_X$.

\begin{definition}
    A \emph{syntactic abstraction} $\mathbf{M}(\alpha_1, \ldots, \alpha_n)$ of arity $n$ for a PPL $\Sigma$ is a document in $\Sigma \cup \set{\alpha_1, \ldots, \alpha_n}$ where $\alpha_1, \ldots, \alpha_n$ are metavariables of some sorts.
    An \emph{instance} $\mathbf{M}(e_1, \ldots, e_n)$ is a document in $\Sigma$ that substitutes  $\alpha_i$ with $e_i$ in $\mathbf{M}(\alpha_1, \ldots, \alpha_n)$ for all $1 \le i \le n$, where $e_i$ and $\alpha_i$ must have a compatible sort.
\end{definition}

\begin{example}\label{ex:syntactic-abstraction}
    \lstinline[language=pp, mathescape]{$\mathbf{M}(\alpha_1, \alpha_2) = \alpha_1$ <> nl <> $\alpha_2$} is a syntactic abstraction for \Lexpressive{}, where $\alpha_1$ and $\alpha_2$ have sort $\mathsf{Doc}$.
    On the other hand, \lstinline[language=pp, mathescape]{$\mathbf{M'}(\alpha_1) =\ $nest $\alpha_1$ nl <> $\alpha_1$} is \textbf{not} a syntactic abstraction because the first occurrence of $\alpha_1$ requires it to have sort $\N$, but the second occurrence requires it to have sort $\DocS$.
    An instance \lstinline[language=pp, mathescape]{$\mathbf{M}($text "a"$,\ $text "b"$)$} is the document \pp{text "a" <> nl <> text "b"}, but \lstinline[language=pp, mathescape]{$\mathbf{M}($text "a"$,\ 1)$} is not an instance due to the incompatible sort.
\end{example}

\begin{definition}
    Let $\Sigma_{\mathsf{base}}$ be a PPL and $\Sigma_{\mathsf{extended}} = \Sigma_{\mathsf{base}} \cup \set{\mathbf{F}}$ where $\mathbf{F}$ has arity $n$ with resulting sort $\DocS$.
    A \emph{syntactic expansion} $\mathsf{expand}_{\mathbf{F}}^{\mathbf{M}}(d)$ from $\Sigma_{\mathsf{extended}}$ to $\Sigma_{\mathsf{base}}$ is a function from $\mathcal{D}_{\mathsf{extended}}$ to $\mathcal{D}_{\mathsf{base}}$ that replaces every occurrence of $\mathbf{F}(e_1, \ldots, e_n)$ with an instance $\mathbf{M}(e_1, \ldots, e_n)$ in $d$, where $\mathbf{F}$ and $\mathbf{M}$ must have compatible arity and sort arguments.
\end{definition}

\begin{example}
    $\mathsf{expand}_{\text{\pp{<$>}}}^{\mathbf{M}}(\cdot)$ is a syntactic expansion from $\Lexpressive{} \cup \set{\text{\;\pp{<$>}}}$ to \Lexpressive{}, where $\mathbf{M}$ is from \Cref{ex:syntactic-abstraction}.
    Hence, $\mathsf{expand}_{\text{\pp{<$>}}}^{\mathbf{M}}(\text{\pp{text "a" <$> text "b"}}) = \text{\pp{text "a" <> nl <> text "b"}}$.
\end{example}

We are now ready to define definability.\footnote{
One important distinction of this definition and Felleisen's counterpart is that PPLs are \emph{total}. 
Hence, observing termination behavior, as done in Felleisen's work, is not feasible in our formulation.}

\begin{definition}
    Let $\Sigma_{\mathsf{base}}$ be a PPL and 
    $\Sigma_{\mathsf{extended}} = \Sigma_{\mathsf{base}} \cup \set{\mathbf{F}}$.
    We say that $\mathbf{F}$ is \emph{definable} by $\Sigma_{\mathsf{base}}$ if there exists a syntactic abstraction $\mathbf{M}$ from $\Sigma_{\mathsf{extended}}$ to $\Sigma_{\mathsf{base}}$ such that for every document $d \in $ $\mathcal{D}_{\mathsf{extended}}$, $\render_{\mathsf{extended}}(d) = \render_{\mathsf{base}}(\mathsf{expand}_{\mathbf{F}}^{\mathbf{M}}(d))$.
\end{definition}

We can now present one of our main results:

\begin{restatable}{theorem}{lexpressiveexpressibility}\label{thm:expressive}
    Every construct in the traditional and arbitrary-choice PPLs is definable in \Lexpressive{}.
\end{restatable}

Despite the result, one might wonder if \Lexpressive{} is actually needed.
Could it be that the arbitrary-choice PPL can already define every construct in the traditional PPL?
As we foreshadowed, the answer to this question is negative. 
However, we must first develop tools that allow us to answer the question, again following the development in Felleisen's work.

\begin{definition}
    A context $C(\alpha)$ for $\Sigma$ is a unary syntactic abstraction for $\Sigma$ where $\alpha$ has sort $\mathsf{Doc}$.
\end{definition}

\begin{definition}
    Given a PPL $\Sigma$ and a relation $R \subseteq 2^{\LayoutT} \times 2^{\LayoutT}$,
    the relation \obsequiv{R}{\Sigma}{d_1}{d_2} holds if and only if $R(\render(C(d_1)), \render(C(d_2)))$ holds for all contexts $C$ in $\Sigma$.\footnote{The relation $E_R^{\Sigma}$ is a generalization of the operational equivalence relation in Felleisen's work.}
\end{definition}

\begin{example}\label{ex:gen-equiv}
    Let $\mathsf{maxWidth} : \LayoutT \to \mathbb{N}$ be a function that computes the maximum length across all lines in the input layout, 
    and lift $\mathsf{maxWidth}$ to work on any set of layouts. 
    That is, $\mathsf{maxWidth}(L) = \set{\mathsf{maxWidth}(l) : l \in L}$.
    Furthermore, let $R = \set{(L_1, L_2) : \mathsf{maxWidth}(L_1) = \mathsf{maxWidth}(L_2)}$.
    \begin{itemize}
    \item \obsequiv{R}{\Lexpressive{}}{\text{\pp{text "a"}}}{\text{\pp{text "b"}}} holds by induction.
    Intuitively, this is because (1) if we only observe the width, the textual content doesn't matter, and (2) there is no construct in \Lexpressive{} that allows us to lay out differently in a way that would affect the width based on the textual content.
    \item On the other hand, \notobsequiv{R}{\Lexpressive{}}{\text{\pp{text "a"}}}{\text{\pp{text "aa"}}}.
    For example, with $C(\alpha) = \alpha$, we have that $\mathsf{maxWidth}(\render_e(C(\text{\pp{text "a"}}))) = \set{1}$, but $\mathsf{maxWidth}(\render_e(C(\text{\pp{text "aa"}}))) = \set{2}$. 
    \end{itemize}
\end{example}

The following theorem provides a tool to prove that a construct is not definable in a PPL.

\begin{restatable}{theorem}{inexpressibility}\label{thm:inexpressibility}
    Given a PPL $\Sigma$ and a construct $\mathbf{F}$, if there exists two documents $d_1$ and $d_2$ in $\Sigma$ and a relation $R$ such that \obsequiv{R}{\Sigma}{d_1}{d_2}, but \notobsequiv{R}{\Sigma \cup \set{\mathbf{F}}}{d_1}{d_2}, 
    then $\mathbf{F}$ is not definable in $\Sigma$.
\end{restatable}

With this tool, we are able to prove that some constructs of $\Sigma_{\mathsf{all}}$ are not definable in the traditional and arbitrary-choice PPLs:

\begin{restatable}{theorem}{inexpressibilityoldlangs}\label{thm:inexpressibility-old-langs}
    The following is true:
    \begin{itemize}
    \item \pp{<>} is not definable in the arbitrary-choice PPL.
    \item \pp{nest} is not definable in the arbitrary-choice PPL.
    \item \pp{group} is not definable in the arbitrary-choice PPL.
    \item \pp{<+>} is not definable in the traditional PPL.
    \end{itemize}
\end{restatable}

Next, we show a relationship between functional completeness and definability.

\begin{restatable}{lemma}{functionalityimpliesinexpressibility}\label{lem:functionality-implies-inexpressibility}
    If $\Sigma$ is not functionally complete, but $\Sigma \cup \set{\mathbf{C}}$ is, then $\mathbf{C}$ is not definable in $\Sigma$.
\end{restatable}

Lastly, we present our final result for this section:  \Lexpressive{} is \emph{minimal} in the sense that each of its constructs is not definable by \Lexpressive{} without it.

\begin{restatable}{theorem}{minimal}\label{thm:minimal}
    For any construct $\mathbf{F}$ of \Lexpressive{}, $\mathbf{F}$ is not definable in $\Lexpressive{} \setminus \set{\mathbf{F}}$.
\end{restatable}

\section{Our printer, \texorpdfstring{\Pexpressive}{Pi\ e}}\label{sec:printer}

In this section, we describe our printer, \Pexpressive, which targets the PPL \Lexpressive{} presented in \Cref{sec:semantics}.
\Pexpressive{} is parameterized by a cost factory and a computation width limit $\cfw$.
We start with an overview of \Pexpressive{}. 
Then, we define a \emph{measure}, which is an output from the core printer that allows us to record a cost and at the same time avoid a full-blown, expensive rendering.
After that, we describe the requirements of the input document structure, which will become important when we analyze the time complexity of the printer.
Then, we present \Pexpressive{}'s printing algorithm, which utilizes the cost factory to achieve optimal and efficient printing.
Finally, we analyze the time complexity of \Pexpressive{}.

\subsection{Overview}

\begin{figure}
  \includegraphics[width=0.95\textwidth]{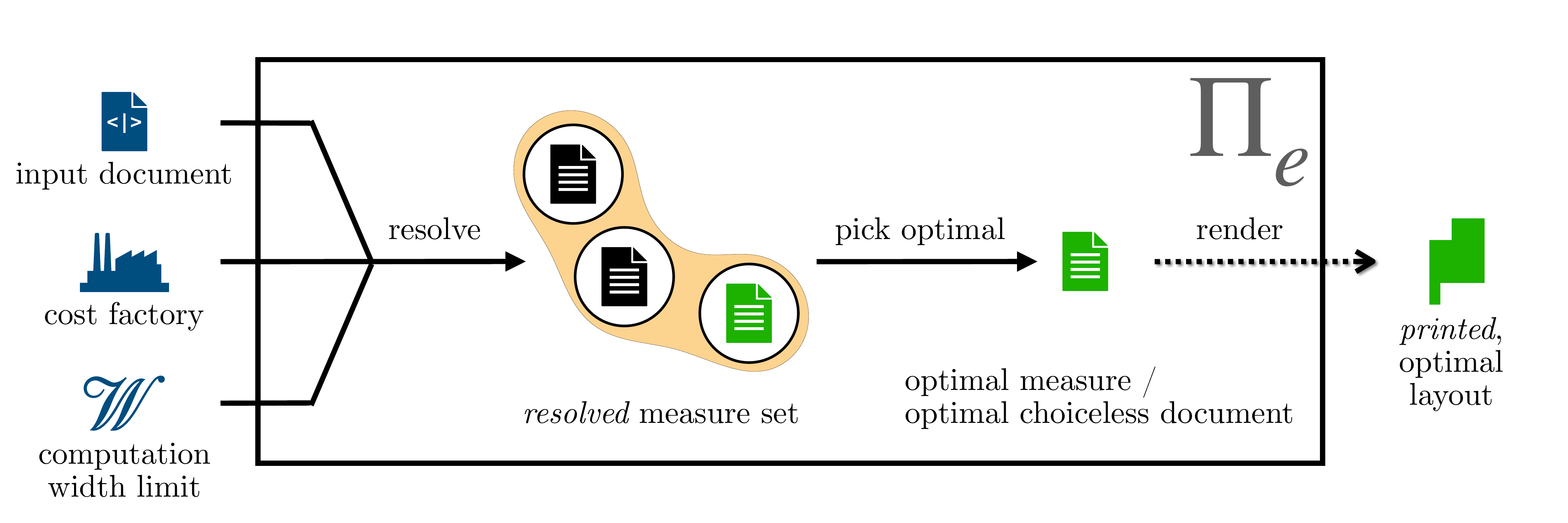}
  \caption{Architecture diagram of our pretty printer, \Pexpressive{}}\label{fig:arch}
\end{figure}

\begin{figure}
  \includegraphics[width=0.95\textwidth]{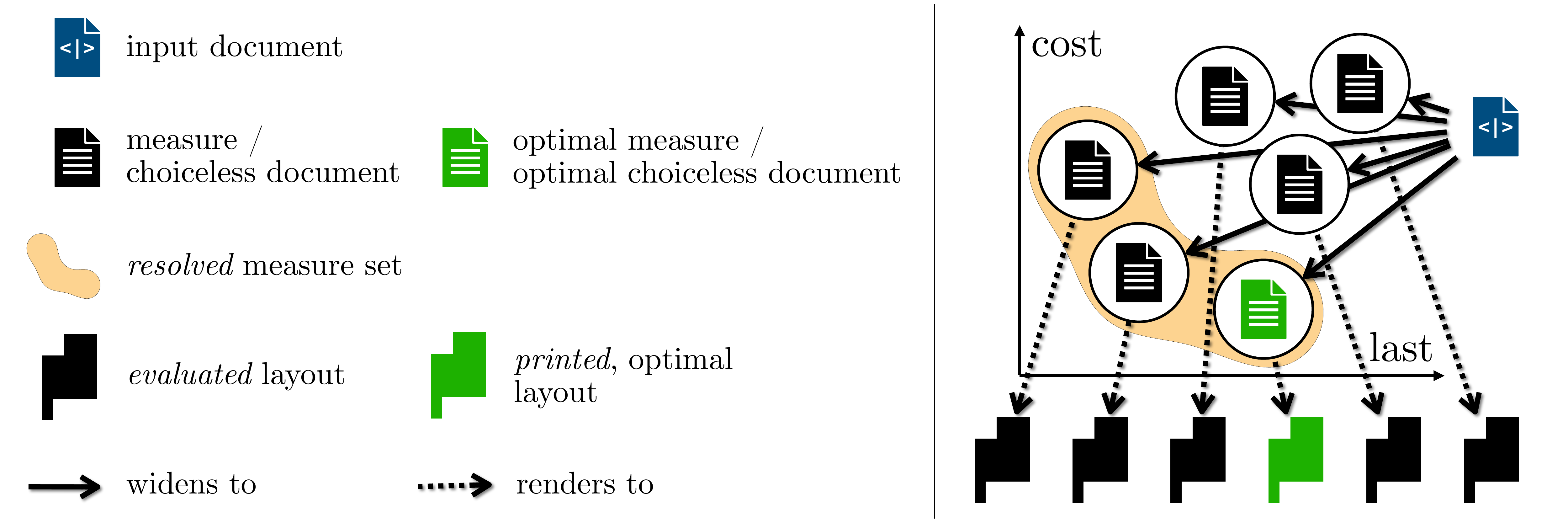}
  \caption{Relationship between evaluation and printing}\label{fig:workflow}
\end{figure}

So far we have defined the \emph{evaluation} of a document, 
which produces the set of possible layouts. 
But when we \emph{print} a document, 
we wish to output only a single, optimal layout.

A na\"ive approach would be to evaluate the input document, via widening and rendering, to all possible layouts, determine costs of these layouts according to a given optimality objective, and then pick one with the least cost as the optimal layout.
However, this approach is not practical for two reasons. 
First, widening could produce exponentially many choiceless documents.
Second, rendering non-optimal choiceless documents is unnecessary and wasteful.

A better approach would utilize early pruning to reduce the search space, and avoid rendering until an optimal choiceless document is first identified.
The need to prune early motivated the design of the cost factory interface shown in \Cref{fig:cost-factory}, which allows \Pexpressive{} to incrementally compute costs to be used for pruning decisions.
Since we wish to avoid full-blown rendering,
we will instead operate on \emph{measures}~\cite{bernardy}, which record the information about a choiceless document required for pruning without expensive rendering.

The workflow of \Pexpressive{} is shown in \Cref{fig:arch}, while \Cref{fig:workflow} shows how it relates to the evaluation of a document.
The printer first resolves choices, with early pruning, to produce a small set of measures that contain the optimal measure.
The set in particular forms a Pareto frontier in the cost and last line length trade-off (\Cref{sec:printer:measure} and \Cref{sec:printer:measure-set}).
We then pick the optimal measure from the set and render its choiceless document to produce an optimal layout.

In the rest of this section, every definition and theorem is implicitly parameterized by a cost factory $\mathcal{F}$ and a computation width limit $\cfw$.

\subsection{Measure}\label{sec:printer:measure}

As presented earlier, 
the resolving phase computes \emph{measure}s. 
Presented in \Cref{fig:measure}, a measure consists of five components: length of last line ($l$), cost ($\mathcal{C}$), choiceless document ($d$), max column position ($\ghosted{x}$), and max indentation ($\ghosted{y}$).
We gray out the last two components because they are ghosted~\cite{ghostvar}: they are only needed for the correctness theorem, and not required in the actual implementation.

\begin{figure}
  \footnotesize

\begin{tabular}{r @{ : } l l r @{ : } l l}

\multicolumn{3}{r}{ Measure $m \in \mT = \meas{l}{\mathcal{C}}{\cl{d}}{x}{y}$ }  \\[1em]

\mlw & $\mT \to \N$ & $\mlw(\meas{l}{\mathcal{C}}{\cl{d}}{x}{y}) = l$ &
\mc & $\mT \to \tau$ & $\mc(\meas{l}{\mathcal{C}}{\cl{d}}{x}{y}) = c$ \\
\mL & $\mT \to \CLDocT$ & $\mL(\meas{l}{\mathcal{C}}{\cl{d}}{x}{y}) = \cl{d}$ \\
\mw & $\mT \to \N$ & $\mw(\meas{l}{\mathcal{C}}{\cl{d}}{x}{y}) = x$ &
\mi & $\mT \to \N$ & $\mi(\meas{l}{\mathcal{C}}{\cl{d}}{x}{y}) = y$ \\[1em]
\end{tabular}

\begin{tabular}{r @{ : } l l }
$\circ$ & $\mT \to \mT \to \mT$ &
$\meas{l_a}{\mathcal{C}_a}{\cl{d}_a}{x_a}{y_a} \circ \meas{l_b}{\mathcal{C}_b}{\cl{d}_b}{x_b}{y_b} =$ \\
\multicolumn{2}{l}{ } & $\quad \meas{l_b}{\mathcal{C}_a \cfcombine \mathcal{C}_b}{ \cl{d}_a\text{ \pp{<>} }\cl{d}_b }{\max(x_a, x_b)}{\max(y_a, y_b)}$ \\[1em]

$\mathsf{adjustNest}$ & $\N \to \mT \to \mT$ & 
$\mathsf{adjustNest}(n, \meas{l}{\mathcal{C}}{\cl{d}}{x}{y}) = \meas{l}{\mathcal{C}}{\text{\pp{nest} } n\ \cl{d}}{x}{y}$ \\

$\mathsf{adjustAlign}$ & $\N \to \mT \to \mT$ & 
$\mathsf{adjustAlign}(i, \meas{l}{\mathcal{C}}{\cl{d}}{x}{y}) = \meas{l}{\mathcal{C}}{\text{\pp{align} } \cl{d}}{x}{\mathsf{max}(y, i)}$ \\[1em]

$\preceq$ & $\mT \to \mT \to \B$ & 
$\meas{l_a}{\mathcal{C}_a}{\cl{d}_a}{x}{y} \preceq \meas{l_b}{\mathcal{C}_b}{\cl{d}_b}{x}{y} = l_a \le l_b \land \mathcal{C}_a \cfle \mathcal{C}_b$ 
\end{tabular}

  \caption{Measure and operations on measures}\label{fig:measure}
\end{figure}

\begin{figure}
  \input{fig/measure-rule}
  \caption{Measure computation from a choiceless document in a printing context}\label{fig:measure-rule}
\end{figure}

\begin{example}
  Let $\cl{d}$ be the choiceless document in \Cref{ex:choiceless-render}.
  With the cost factory in \Cref{ex:cost-factory-linear} and $w = 6$,
  the choiceless document is rendered at the column position $3$ 
  and indentation level $0$ to the second layout in \Cref{fig:cost-factory-illustration}, 
  with the cost $(8, 3)$.
  The column position of the last line is 1.
  The maximum column position attained is 10 (on the first line), 
  and the maximum indentation level attained is 2.
  Thus, the computed measure is $\meas{1}{(8, 3)}{\cl{d}}{10}{2}$.
\end{example}

\Cref{fig:measure-rule} shows rules that define measure computation.
The judgment $\measL{d}{c}{i}{m}$ states that when we compute the measure of $\cl{d} \in \CLDocT$ placed at the column position $c \in \N$ with indentation level $i \in \N$, the resulting measure is $m \in \mT$.
To simplify the core printer, we (temporarily) remove \pp{flatten} from \Lexpressive{}.
This allows us to eliminate the flattening mode parameter, which implicitly defaults to $\bot$.
Toward the end of this section, 
we will show how to add support for \pp{flatten} back.

The rules are largely standard.
They reflect the actual rendering defined by $\Downarrow_{\mathcal{R}}$, and utilize the cost factory in a straightforward way.
The rules use a helper operator function $\circ$ to concatenate two measures, and helper functions $\mathsf{adjustNest}$ and $\mathsf{adjustAlign}$ to construct a correct measure for \pp{nest} and \pp{align}. 
These functions are defined in \Cref{fig:measure}.
Notably, the \textsc{LineM} rule creates a measure whose $\mathsf{maxc}$ is $\mathsf{max}(c, i)$ because before placing the newline, the column position is $c$, and after placing the newline, the column position is $i$.
The \textsc{AlignM} rule creates a measure whose $\mathsf{maxi}$ is $\mathsf{max}(y, i)$ where $y$ is obtained via the recursive computation. 
This is because the recursive computation discards the current indentation level, so we need to specifically record the information. 

$\Downarrow_{\mathbb{M}}$ is deterministic and total.
It is also correct with respect to $\Downarrow_{\mathcal{R}}$.

\begin{theorem}\label{thm:measure-correctness}
  For any $\cl{d} \in \CLDocT$ and $c, i \in \N$,
  there exists a maximum indentation $y$ such that
  \begin{itemize}
    \item if $\renderL{\cl{d}}{c}{i}{\bot}{[s]}$, then
    $\measL{\cl{d}}{c}{i}{\meas{c + |s|}{\textsc{Cost}(c, [s])}{\cl{d}}{c + |s|}{y}}$.
    \item if $\renderL{\cl{d}}{c}{i}{\bot}{[s, s_1, \ldots, s_n, t]}$, then\\
    $\measL{\cl{d}}{c}{i}{\meas{|t|}{\textsc{Cost}(c, [s, s_1, \ldots, s_n, t])}{\cl{d}}{\max(c + |s|, |s_1|, \ldots, |s_n|, |t|)}{y}}$
  \end{itemize}
\end{theorem}

So far, we have only considered the measure computation for a choiceless document. 
When we take the choice operator into account, there could be multiple measures under the same printing context.
The main operation that we can perform on these measures is finding domination $\preceq$, also presented in \Cref{fig:measure}.
$m_a \preceq m_b$ when both the cost and the last length of $m_a$ are no worse than those of $m_b$.
The fact that $m_a \preceq m_b$ is useful because it allows us to prune $m_b$ away immediately.

\subsection{Measure Set}\label{sec:printer:measure-set}

\begin{figure}
  \footnotesize

{\setlength{\tabcolsep}{3pt}
\begin{tabular}{l r l l}
    Measure set $S \in \msT$ & $\defn$ & $\mstainted(\hat{m})$ & where $\hat{m}$ is a promise that can be forced to a measure \\ 
    & $\vbar$ & $\msset([m_1, \ldots^+, m_n])$ & where $\mlw(m_1) > \ldots > \mlw(m_n)$ and $\forall i \ne j, \neg (m_i \preceq m_j \lor m_j \preceq m_i)$\\[1em]
\end{tabular}}

\begin{tabular}{r @{ : } l l l}
$\mathsf{taint}$ & $\msT \to \msT$ & $\mathsf{taint}(\mstainted(m)) = \mstainted(m)$\\
\multicolumn{2}{l}{} & $\mathsf{taint}(\msset([m_0, m_1, \ldots, m_n])) = \mstainted(m_0)$ \\[1em]

$\mathsf{lift}$ & $\msT\!\to\!(\mT\!\to\!\mT)\!\to\!\msT$ & $\mathsf{lift}(\mstainted(m), f) = \mstainted(f(m))$\\
\multicolumn{2}{l}{} & $\mathsf{lift}(\msset([m_1, \ldots^+, m_n]), f) = \msset([f(m_1), \ldots^+, f(m_n)])$\\[1em]

$\mathsf{dedup}$ & $\overrightarrow{\mT} \to \overrightarrow{\mT}$ & $\mathsf{dedup}([m, m', m_1, \ldots, m_n]) = \mathsf{dedup}([m', m_1, \ldots, m_n])$ & if $m' \preceq m$\\
\multicolumn{2}{l}{} & $\mathsf{dedup}([m, m', m_1, \ldots, m_n]) = [m] @ \mathsf{dedup}([m', m_1, \ldots, m_n])$ & if $m' \not\preceq m$\\
\multicolumn{2}{l}{} & $\mathsf{dedup}([m]) = [m]$\\[1em]

$\uplus$ & $\msT \to \msT \to \msT$ & $S \uplus \mstainted(m) = S$\\
\multicolumn{2}{l}{} & $\mstainted(m) \uplus \msset([m_1, \ldots^+, m_n]) = \msset([m_1, \ldots^+, m_n])$\\
\multicolumn{2}{l}{} & 
\multicolumn{2}{l}{$\msset([m_1, \ldots^+, m_n]) \!\uplus\!\msset([m'_1, \ldots^+, m'_{n'}]) = \msset([m_1, \ldots^+, m_n]\!\uplus\![m'_1, \ldots^+, m'_{n'}])$} \\[1em]

$\uplus$ & $\overrightarrow{\mT} \to \overrightarrow{\mT} \to \overrightarrow{\mT}$ & 
$[] \uplus [m_1, \ldots^+, m_n] = [m_1, \ldots^+, m_n]$ \\
\multicolumn{2}{l}{} & $[m_1, \ldots^+, m_n] \uplus [] = [m_1, \ldots^+, m_n]$ \\
\multicolumn{4}{l}{$[m_0, m_1, \ldots, m_n] \uplus [m'_0, m'_1, \ldots, m'_{n'}] = \begin{cases}
    [m_0, m_1, \ldots, m_n] \uplus [m'_1, \ldots, m'_{n'}]  & \text{ if } m_0 \preceq m'_0 \\
    [m_1, \ldots, m_n] \uplus [m'_0, m'_1, \ldots, m'_{n'}]  & \text{ if } m'_0 \preceq m_0 \\
    [m_0] @ ([m_1, \ldots, m_n] \uplus [m'_0, m'_1, \ldots, m'_{n'}])  & \text{ if } \mlw(m_0) > \mlw(m'_0) \\
    [m'_0] @ ([m_0, m_1, \ldots, m_n] \uplus [m'_1, \ldots, m'_{n'}])  & \text{ otherwise } 
\end{cases}$}
\end{tabular}
  \caption{Measure set and the merge operation on measure sets. 
  $@$ denotes a list concatenation.
  We treat a promise $\hat{m}$ and a measure $m$ interchangeably, as they can be straightforwardly casted to each other.}\label{fig:measure-set}
\end{figure}

Resolving a document (in a printing context)  produces a small set of measures.
To accommodate taintedness mentioned in \Cref{sec:tour:limit},
\Cref{fig:measure-set} defines a measure set to be either a non-empty \msset{} of untainted measures where no measure dominates the other, or a \mstainted{} singleton set of a promise $\hat{m}$ that can be forced to a measure.
The \msset{}, by definition, forms a Pareto frontier.
To aid computation, we represent the \msset{} with a list ordered by the cost in strict ascending order (and therefore the last length in strict descending order).
We are able to do so because in a Pareto frontier, all \mlw{} and \mc{} values must be distinct.

The main operation that we can perform on measure sets is merging two measure sets ($\uplus$), shown in \Cref{fig:measure-set}, where we prefer a \msset{} over a \mstainted{}.
The merge operation maintains the Pareto frontier invariant, by doing the merge in the style of the merge operation in merge sort, although the Pareto frontier merging can also prune measures away during the operation.
One important ``quirk'' of this merge operation is that it is \emph{left-biased} in the presence of taintedness.
If two tainted measure sets are merged, the result is always the left one.
This means the order of arguments to the merge operation is important, as we will see in the next subsections.

Other operations on measure sets which are used in subsequent sections are $\mathsf{taint}$, $\mathsf{lift}$, and $\mathsf{dedup}$. 
$\mathsf{taint}$ taints a measure set. 
When tainting a \msset{}, we choose to pick the first measure from the \msset{} because it has the least cost, which is a greedy heuristic.
$\mathsf{lift}$ adjusts measures in a measure set.
Lastly, $\mathsf{dedup}$ prunes measures 
that are sorted by $\mlw{}$ in strictly decreasing order and by $\mc{}$ in non-strictly increasing order, 
so that the result conforms to the Pareto frontier invariant.

\subsection{The Document Structure}\label{sec:printer:doc-structure}

\Cref{sec:related:arbitrary} showed that we need to handle document sharing by treating the input document as a DAG.
However, documents cannot be arbitrarily shared, as the following example shows:

\begin{example}
  The following document $\mathsf{mk}(n)$ has a DAG size of $O(n)$.
  However, resolving it necessitates $O(2^n)$ units of computation, as the printing contexts are all different. 
  This is bad news because it means that resolving could take time exponential in the input size.
\begin{lstlisting}[language=pp]
let rec mk (n : int): doc = 
  if n = 0 then text "x" 
  else let shared = mk (n - 1) in shared <> shared
\end{lstlisting}
\end{example}
However, we argue that the above document is not \emph{properly shared}, because the sub-documents are not shared \emph{across choices}, which is how sharing is employed in practice. 
The corresponding properly shared document would have $O(2^n)$ DAG size, so $O(2^n)$ units of computation are still linear in the input size.
To make this precise, we provide the following definitions:

\begin{definition}
  Given a document $d \in \mathcal{D}_e$, 
  $G(d)$ is a DAG rooted at $d$ whose edge in the graph connects a document to its direct subdocuments.
\end{definition}

\begin{definition}
  A document $d \in \mathcal{D}_e$ is \emph{properly shared} if for any two vertices $d_a$ and $d_b$ in $G(d)$, if $p_1$ and $p_2$ are two distinct paths from $d_a$ to $d_b$, then there exists a common document $d'$ such that (1) $d'$ is a \pp{<|>}; (2) $d'$ occurs in both $p_1$ and $p_2$; and (3) $d'$ is not $d_b$.
\end{definition}

\Cref{fig:sharing:dag} shows a properly shared document (assuming that $D$ is properly shared).
It illustrates two paths where $d_a$ is the root node, $d_b$ is $D$, and $d'$ is $d_a$.
In practice, non-properly shared documents can still be processed by \Pexpressive{}, and in fact can even make resolving faster when a shared document is resolved under the same printing context.
However, this shared document would be effectively duplicated when it is resolved in different contexts.
For simplicity, we only consider properly shared documents as the input to \Pexpressive{} in this paper.

\subsection{The Resolver}

\begin{figure}
  \input{fig/printer}
  \caption{The resolver}\label{fig:printer}
\end{figure}

We now formally define the core of \Pexpressive{}, which is the resolver. 
It is described in \Cref{fig:printer}, which is a fusion of widening in \Cref{fig:document-semantics} and measure computation in \Cref{fig:measure-rule},
with early pruning inherent in the merge operation and extra bookkeeping for taintedness.
The judgment $\pL{d}{c}{i}{S}$ states that a properly shared document $d \in \mathcal{D}_e$ at a column position $c \in \N$ with an indentation level $i \in \N$ resolves to a measure set $S$.

\paragraph{Resolving Text} 
If placing the text would exceed \cfw{} or the indentation level is beyond \cfw{}, 
the \textsc{TextRSTnt} rule returns a \mstainted{}.
Otherwise, the \textsc{TextRS} rule returns a singleton \msset{}.

\paragraph{Resolving Newlines}
Resolving a \pp{nl} is similar to resolving a \pp{text}, 
but we only need to consider the current column position and indentation level, as resolving the newline does not change the column position.
The \textsc{LineRSTnt} and \textsc{LineRS} rules cover these two cases.

\paragraph{Resolving Nesting}
Resolving a \pp{nest} is handled by the \textsc{NestRS} rule, which recursively resolves its sub-document with the indentation level changed.
The recursive resolving determines whether the measure set will be a \msset{} or \mstainted{}.
In all cases, the result is adjusted to construct correct choiceless documents. 

\paragraph{Resolving Alignment}
Resolving an \pp{align} is similar to resolving \pp{nest}.
However, because the recursive resolving discards the current indentation level, which could exceed \cfw{}, 
we need to taint the measure set when the indentation level is beyond \cfw{}.
The \textsc{AlignRSTnt} rule handles such cases, 
and the \textsc{AlignRS} rule handles the other possibilities.

\paragraph{Resolving Choices}
The \textsc{UnionRS} rule recursively resolves its two sub-documents and then merges the resulting measure sets.
As mentioned in \Cref{sec:printer:measure-set}, 
the merge operation is left-biased.
Therefore, the left sub-document will be preferred over the right sub-document if exceeding \cfw{} is unavoidable.
It is possible to employ a heuristic to remove this bias, as discussed in \appendixref{sec:discussion}.

\paragraph{Resolving Unaligned Concatenation}

Resolving a \pp{<>} is done through the \textsc{ConcatRSTnt} and \textsc{ConcatRS} rules, which handle the two possibilities of measure set types obtained from the left sub-document's recursive resolving.
Notably, the \textsc{ConcatRS} rule employs $\Downarrow_{\mathbb{RSC}}$ to help us concatenate a left measure from the left measure set with a right measure set.

\vspace{2mm}

$\Downarrow_{\mathbb{RS}}$ is deterministic and total.
This allows us to define the top-level printer as $\Pexpressive(d) = l$ where $\pL{d}{0}{0}{[m_0, m_1, \ldots, m_n]}$ and $\renderL{\mL(m_0)}{0}{0}{\bot}{l}$, which consumes a properly shared document $d$, resolves it to a set of measures, picks the measure with the least cost, and renders the associated choiceless document to produce a layout. (Our implementation further fuses resolving and rendering together, as described in \appendixref{sec:discussion}.)

While the rules above are enough for correctness, implementing these rules requires further consideration. 
As we will see in \Cref{lem:print-beyond-limit}, any resolving beyond \cfw{} would eventually result in a tainted measure set. 
Hence, \Pexpressive{} should \emph{immediately} delay the computation for any resolving beyond \cfw{}.
\Pexpressive{} should also \emph{memoize} the computation, so that on identical documents and printing contexts within \cfw{}, the result of the previous computation is reused.

We claim that $\Pexpressive(d)$ consumes a properly shared document $d$ in \Lexpressive{} and produces an optimal layout among $\render_e(d)$ within \cfw{}.
We prove this claim in the next subsection.

\subsection{Correctness of \texorpdfstring{\Pexpressive}{Pi\ e}}

$\Downarrow_{\mathbb{RS}}$ is correct with respect to $\Downarrow_{\mathbb{M}}$.
Two theorems govern the correctness.
The first theorem states that the core printer returns a measure set that contains a measure that is no worse than any measure within the computation width limit from all possible measures.

\begin{theorem}[Optimality]
For any $d \in \mathcal{D}_e$, $c \in \N$, $i \in \N$, if the following conditions hold
\begin{multicols}{3}
\begin{itemize}
  \item $\pL{d}{c}{i}{S}$
  \item $\widen{d}{\cl{D}}$
\end{itemize}
\columnbreak
\begin{itemize}
  \item $\cl{d} \in \cl{D}$
  \item $\measL{\cl{d}}{c}{i}{m}$
\end{itemize}
\columnbreak 
\begin{itemize}
  \item $\mw(m) \le \cfw$
  \item $\mi(m) \le \cfw$
\end{itemize}
\end{multicols}
then $S = \msset([m_1, \ldots^+, m_n])$. 
Furthermore, there exists $i$ such that $m_i \preceq m$.
\end{theorem}

The second theorem states that measures in the resulting measure set are valid.

\begin{theorem}[Validity]
For any $d \in \mathcal{D}_e$, $c \in \N$, $i \in \N$ with $\widen{d}{\cl{D}}$, if $\pL{d}{c}{i}{\msset([m_1, \ldots^+, m_n])}$, then for each $i$, there exists $\cl{d}$ such that $\cl{d} \in \cl{D}$ and $\measL{\cl{d}}{c}{i}{m_i}$.
Likewise, if $\pL{d}{c}{i}{\mstainted(m_0)}$, then there exists $\cl{d}$ such that $\cl{d} \in \cl{D}$ and $\measL{\cl{d}}{c}{i}{m_0}$.
\end{theorem}

The correctness of \Pexpressive{} follows immediately.

While the above theorems guarantee the correctness of the result that the printer produces, they do not guarantee efficiency.
The following lemmas provide some properties of the printer that allow us to reason about its efficiency.

\begin{lemma}\label{lem:number-of-measures}
For any $d \in \mathcal{D}_e$, $c \le \cfw$, $i \le \cfw$, if $\pL{d}{c}{i}{\msset([m_1, \ldots, m_n])}$, then $n \le \cfw + 1$.
\end{lemma}

\begin{lemma}\label{lem:print-beyond-limit}
For any $d \in \mathcal{D}_e$, if $c > \cfw$ or $i > \cfw$ and $\pL{d}{c}{i}{S}$, then $S$ is a \mstainted{}.
\end{lemma}

We now informally prove the efficiency of \Pexpressive{} that we claimed in \Cref{sec:intro}.
The proof sketches are provided in \appendixref{sec:proofs}.

\begin{restatable}{theorem}{timecomplexity}
  The time complexity of \Pexpressive{} is $O(n \cfw^4)$ where $n$ is the DAG size of the document.
\end{restatable}

\begin{restatable}{theorem}{timecomplexityarbfragment}
  If a document $d$ is in the arbitrary-choice PPL, \Pexpressive{} can print $d$ in $O(n \cfw^3)$.
\end{restatable}

\subsection{Handling Flattening}\label{sec:printer:flatten}

To support \pp{flatten}, we make it a function that walks its sub-document and replaces all \pp{nl} with \pp{text " "}.
The walk is memoized and preserves the original identity of the document whenever possible (i.e. if nothing is flattened in sub-documents, then the document itself is returned unchanged without creating a new document).
Thus, each document can be flattened at most once. 
This flattening creates at most $O(n)$ new documents without destroying the shared structure in the original document.  
We therefore achieve the functionality of \pp{flatten} without affecting the time complexity of the printer.

\section{Implementation}\label{sec:impl}

We implement \Pexpressive{} in OCaml and Racket.
The printer, which we call \prettiester{}, is further refined to be more efficient and practical. 
\prettiester{} also includes more practical constructs that do not fit well to the formalism in this paper. 
We describe these refinements and additional constructs in \appendixref{sec:discussion}.
The OCaml \prettiester{}, as a reference implementation, is used for comparing against other printers in \Cref{sec:eval}. 
The Racket \prettiester{} has more features, and it has been used to implement a code formatter for the Racket programming language.

In these implementations, we extend the cost factory interface in \Cref{fig:cost-factory} so that $\cfnl$ is now a procedure that takes an indentation level $i$ as an input, and returns the cost of a newline along with $i$ indentation spaces, with a contract that $\forall i, i' \in \N.\ i \le i' \to \cfnl(i) \cfle \cfnl(i')$.
That is, $\cfnl(i) = \cfnl \cfcombine \cftext(0, i)$ was not customizable before, but it is now customizable.\footnote{This change requires adjustments to many definitions and theorems, and we have done so for our Lean formalization. For example, to make \Cref{thm:measure-correctness} hold, we need to keep indentation spaces in the definition of layouts (\Cref{sec:layout}).}
\prettiester{} then provides a pre-defined cost factory that is like \Cref{ex:cost-factory-quadratic}, but with $\cfnl(i) = (0, 1)$.
\section{Evaluation}\label{sec:eval}

This section evaluates the performance and optimality of \prettiester{}.
The evaluation consists of two parts.
First, we compare \prettiester{} against Wadler/\citet{leijen} and \citet{bernardy:implementation}'s printers, which are popular practical printers with capabilities from the traditional and arbitrary-choice PPLs.
Second, we evaluate the Racket code formatter, which uses \prettiester{} as its foundation.
The evaluation aims to answer the following questions:
\begin{enumerate}
\item Does \prettiester{} run fast in practice?
\item Does \prettiester{} produce pretty layouts in practice?
\end{enumerate}
All experiments are performed on an Apple M2 MacBook Pro with 16GB of RAM.
We describe the experiments and benchmarks in \Cref{sec:eval:comparison} and \Cref{sec:eval:fmt}, and discuss the results in \Cref{sec:eval:results}.

\subsection{Comparison of Printers}\label{sec:eval:comparison}

We compare OCaml \prettiester{} against the latest version (1.2.1) of Wadler/Leijen's printer, and the ``camera ready version'' of Bernardy's printer\footnote{We also tried other versions of Bernardy's printer, such as the commit 006fa0e8, which is the version right before the \pp{<|>} operator was removed, and supposedly more optimized than the camera ready version. 
Unfortunately, we find that it has a severe performance deficiency. 
When attempting to replicate the experiments in \citet{bernardy}, we find that formatting the 10k-line-JSON file takes about 80 seconds, which is much slower than the 145 milliseconds reported in the paper.}.
This ``camera ready version'' consists of two printers: the ``\naive{}'' variant, which is presented in the paper, and the ``practical'' implementation, which has more features (such as unavoidable overflow handling) but suffers from exponential time complexity when the DAG structure unfolds, as discussed in \Cref{sec:related}.
We manually remove the capability to customize the width limit from the latter to avoid the issue.
Both variants are used for the evaluation, since the \naive{} variant does not have necessary features for some benchmarks.

\prettiester{} is instantiated with the cost factory in \Cref{sec:impl}, 
with a page width limit of 80 (unless indicated otherwise).
We run \prettiester{} twice with different computation width limits (once with $\cfw = 100$, unless indicated otherwise, and once with $\cfw = 1000$), in order to observe the effect of the tainting system and how it affects the performance.

\begin{table}[t]
\scriptsize
\caption{Comparison between \prettiester{} in different configurations and other printers. 
For each printer and configuration, the first column reports the running time, and the second column reports the line count of the output layout.
\prettiester{} has an additional third column, where \cmark{} indicates that the rendering to the output layout fits \cfw{} and \xmark{} indicates that the rendering to the output layout is tainted.
``N/A'' means the benchmark is not applicable. 
\timeout~indicates that running the benchmark exceeds the timeout of 60 seconds.
``-'' means the data is not collected.
A grayed row indicates an output mismatch among the printers/configurations. 
The bolded line count signals that in our manual inspection, the associated layout is the prettiest.
}\label{table:perf}
\begin{tabular}{ l|rrc|rrc|rr|rr|rr }
\toprule
\multirow{2}{*}{Benchmark} & \multicolumn{6}{c|}{\prettiester{}} & \multicolumn{2}{c|}{\multirow{2}{*}{Wadler/Leijen}} & \multicolumn{4}{c}{Bernardy} 
\\ 
& \multicolumn{3}{c|}{default \cfw{} (usually 100)} & \multicolumn{3}{c|}{$\cfw = 1000$} & & & \multicolumn{2}{c|}{Na\"ive} & \multicolumn{2}{c}{Practical} \\
\midrule
Concat10k & 
0.000 s & 1 & \xmark & 
0.000 s & 1 & \xmark & 
0.002 s & 1 & 
N/A & - & 
0.433 s & 1\\
Concat50k & 
0.002 s & 1 & \xmark & 
0.002 s & 1 & \xmark & 
0.011 s & 1 & 
N/A & - & 
14.626 s & 1\\
FillSep5k & 
0.010 s & 668 & \cmark & 
0.010 s & 668 & \cmark & 
0.004 s & 668 & 
3.097 s & 668 & 
\timeout & -\\
FillSep50k & 
0.190 s & 6834 & \cmark & 
0.190 s & 6834 & \cmark & 
0.035 s & 6834 & 
\timeout & - & 
\timeout & -\\
Flatten8k & 
0.018 s & 7986 & \cmark & 
0.016 s & 7986 & \cmark & 
3.346 s & 7986 & 
N/A & - & 
N/A & -\\
Flatten16k & 
0.036 s & 15986 & \cmark & 
0.037 s & 15986 & \cmark & 
18.816 s & 15986 & 
N/A & - & 
N/A & -\\
SExpFull15 & 
3.027 s & 4107 & \cmark & 
5.437 s & 4107 & \cmark & 
0.045 s & 4107 & 
0.647 s & 4107 & 
0.911 s & 4107\\
SExpFull16 & 
5.255 s & 8246 & \cmark & 
14.232 s & 8246 & \cmark & 
0.091 s & 8246 & 
1.251 s & 8246 & 
1.802 s & 8246\\
\rowcolor{gray!30}
RandFit1k & 
0.100 s & \textbf{629} & \cmark & 
0.229 s & \textbf{629} & \cmark & 
0.003 s & 943 & 
0.048 s & \textbf{629} & 
0.074 s & \textbf{629}\\
\rowcolor{gray!30}
RandFit10k & 
1.047 s & \textbf{7861} & \cmark & 
4.420 s & \textbf{7861} & \cmark & 
0.037 s & 10459 & 
0.534 s & \textbf{7861} & 
0.855 s & \textbf{7861}\\
\rowcolor{gray!30}
RandOver1k & 
0.058 s & \textbf{1531} & \xmark & 
0.904 s & \textbf{1531} & \cmark & 
0.005 s & 1635 & 
N/A & - & 
0.065 s & 1105\\
\rowcolor{gray!30}
RandOver10k & 
0.405 s & \textbf{15027} & \xmark & 
16.553 s & \textbf{15027} & \cmark & 
0.108 s & 16015 & 
N/A & - & 
1.103 s & 7953\\
JSON1k & 
0.001 s & 564 & \cmark & 
0.001 s & 564 & \cmark & 
0.003 s & 564 & 
N/A & - & 
0.005 s & 564\\
JSON10k & 
0.007 s & 5712 & \cmark & 
0.007 s & 5712 & \cmark & 
0.018 s & 5712 & 
N/A & - & 
0.097 s & 5712\\
\rowcolor{gray!30}
JSONW & 
0.001 s & \textbf{721} & \xmark & 
0.001 s & \textbf{721} & \cmark & 
0.002 s & \textbf{721} & 
N/A & - & 
0.005 s & 709\\
\bottomrule
\end{tabular}
\end{table}

\begin{table}[t]
\scriptsize
\caption{The code formatter benchmarks. 
The table is in the same format as the \prettiester{} column in \Cref{table:perf}.}\label{table:racket}
\begin{tabular}{ l|rrc|rrc }
\toprule
Benchmark & \multicolumn{3}{c|}{$\cfw = 100$} & \multicolumn{3}{c}{$\cfw = 1000$} \\
\midrule
\rowcolor{gray!30}
\texttt{class-internal} & 0.325 s & 5750 & \xmark & 0.307 s & \textbf{5751}  & \cmark\\
\texttt{xform} & 0.372 s & 5154 & \xmark & 0.417 s & 5154  & \cmark\\
\bottomrule
\end{tabular} \rulesep 
\begin{tabular}{ l|rrc|rrc }
\toprule
Benchmark & \multicolumn{3}{c|}{$\cfw = 100$} & \multicolumn{3}{c}{$\cfw = 1000$} \\
\midrule
\texttt{list} & 0.025 s & 993 & \cmark & 0.025 s & 993  & \cmark\\
\texttt{hash} & 0.020 s & 83 & \cmark & 0.020 s & 83  & \cmark\\
\bottomrule
\end{tabular}

\end{table}

The benchmarks (\Cref{table:perf}) are mostly taken from \citet{bernardy}, and we add a few more to test basic constructs. 
While Leijen's printer is expressive enough to handle all benchmarks (due to the inclusion of \pp{align} to support aligned concatenation in addition to constructs from the traditional PPL), 
Bernardy's printers are not applicable to benchmarks that require constructs from the traditional PPL.
Furthermore, Bernardy's \naive{} printer is not applicable to benchmarks that require extra features like unavoidable overflow handling.

In more detail, the benchmarks test the following kinds of documents:

\begin{description}

\item [Concat] benchmarks test a long chain of concatenations, which are identified by \citet{peyton} as a source of quadratic time complexity in Hughes' printer.

\item [FillSep] benchmarks test the \texttt{fillSep} construct (also known as \texttt{fill}), which performs word wrapping. 

\item [Flatten] benchmarks test repeated flattening, as shown in \appendixref{fig:wadler-time-complexity} in \appendixref{sec:appendix-survey}.

\item [SExpFull] benchmarks are the last two data points from the ``full tree'' benchmark in \citet{bernardy}'s paper.
They create complete binary trees and print them as S-expressions.

\item[RandFit] benchmarks \citep{bernardy} are similar to SExpFull, but use random Dyck paths to generate random trees and filter only those that fit within the page width limit.

\item [RandOver] benchmarks are like RandFit with the opposite filtering.

\item [JSON] benchmarks are also from \citet{bernardy}'s paper.
They format large JSON files.
\item [JSONW] benchmark is the same as JSON1k but with a page width limit of 50 instead of 80, and 
we further adjust \prettiester{}'s default \cfw{} from 100 to 60 to test the tainting system.
\end{description}

\subsection{Racket Code Formatter}\label{sec:eval:fmt}

We evaluate the effectiveness of a Racket code formatter that uses the Racket \prettiester{} as its foundation.
Racket~\cite{racket} is a programmable programming language.
Its main syntax is S-expression based, but this can be customized via its \lstinline[language=literal]{#lang} protocol to read an arbitrary syntax.
Even in the S-expression syntax, users can define custom forms via the macro system.
Our long-term plan for the code formatter is to make it extensible to support any syntax and custom forms.
\prettiester{} is thus a natural choice as a foundational printer, due to its expressiveness.

The code formatter currently supports only S-expression formatting. 
However, the task is already challenging.
While the S-expression syntax may look simple and uniform, 
Racket users employ a variety of styles for different forms to make them look distinctive in order to improve readability.
Each function application, for example, has three possible styles (while most languages have two function application styles).
The search space of the code formatter is thus quite large.

The benchmarks (\Cref{table:racket}) consist of files of different sizes from the Racket language codebase\footnote{\url{https://github.com/racket/racket/tree/master/racket/collects} at commit 4f1a2bd4}. 
\texttt{class-internal} and \texttt{xform} are the two largest files.
We use the code formatter to format these files with the page width limit of 80. 
We run the code formatter twice, once with $\cfw = 100$ and once with $\cfw = 1000$.

\subsection{Results}\label{sec:eval:results}

\paragraph{Performance}
The benchmarking results in \Cref{table:perf} and \Cref{table:racket} show that overall, \prettiester{} is sufficiently fast in practice.
While not the fastest, 
it can process large, practical workloads \texttt{class-internal} and \texttt{xform} under a second.
Furthermore, it provides a performance guarantee even on tricky inputs.
The same is not true for other printers.
The Flatten benchmarks work very poorly for Wadler's printer, and the FillSep benchmarks work very poorly for Bernardy's printer.
Interestingly, Bernardy's \naive{} printer is faster than its practical variant, even though the latter is more optimized; this is due to the extra features that the practical printer needs to support.
\prettiester{}, by contrast, is set to support these features from the start.

We note two interesting observations of \prettiester{}. 
First, it performs poorly on SExpFull relative to other printers.
This is due to the memory pressure from memoization.
Better engineering effort may be able to alleviate this issue.
Second, although the time complexity of \Pexpressive{} is $O(n\cfw^4)$, this worst case behavior happens only if Pareto frontiers are always full. 
In practice, this is not the case\footnote{This observation also applies to Bernardy's printers, which are also based on Pareto frontiers.}, as evidenced by the fact that increasing \cfw{} tenfold does not multiply the running time by $10^4$.
On the contrary, increasing \cfw{} does not affect the running time at all on most benchmarks.

\paragraph{Optimality}\label{sec:eval:optimality}
We find that \prettiester{} is the prettiest compared to others, offering high quality output when we use the cost factory described in \Cref{sec:impl}.
\Cref{table:perf} shows (via line count) that the output layouts in many benchmarks agree in all printers.
The exceptions are RandFit, RandOver, and JSONW benchmarks.
Upon manual inspection, we find that the layouts produced by \prettiester{} are better.
JSONW and RandOver are cases where there is an unavoidable overflow, causing Bernardy's printer to overflow more than necessary.
\appendixref{fig:bernardy-prettiness} in \appendixref{sec:appendix-survey} demonstrates this problem.
RandFit and RandOver are cases where the greedy minimization and the \pp{align} construct in Leijen's printer interact poorly, as discussed in \citet{bernardy}'s paper.

It should also be noted that neither Leijen's nor Bernardy's printers support custom optimality objectives, 
as their optimality objectives are integral to their algorithms.
\prettiester{}, by contrast, allows users to customize optimality objective via the cost factory.

Lastly, we evaluate the effectiveness of the tainting system.
For almost every benchmark that has a tainted rendering (\xmark) with the default \cfw{}, we find that using $\cfw = 1000$ in an attempt to avoid taintedness\footnote{Therefore, the Concat benchmarks do not count, since they are still tainted afterwards. The benchmarks are not interesting anyway, since there is no choice in the documents, so the output layouts are always optimal.} yields the same result, confirming the optimality of the output layout. 
The only exception is the \texttt{class-internal} benchmark in \Cref{table:racket}, for which the output layouts are different in one line and otherwise identical, because the greedy heuristic in the $\mathsf{taint}$ operation prunes the optimal choice away.
This demonstrates that despite being tainted, and thus no longer guaranteed to be optimal, the output layout is still reasonable (at least with respect to the cost factory that we employ and the heuristic to avoid bias described in \appendixref{sec:discussion}).

\section{Conclusion}\label{sec:conclusion}

We have described \Pexpressive{}, an expressive printer that supports a variety of optimality objectives and is practically efficient.
We developed a framework for reasoning about the expressiveness of PPLs, and we used this framework to guide the design of the PPL that \Pexpressive{} targets.
By surveying existing pretty printers, we have shown that \Pexpressive{} is well-placed in the design space of printers.
\Pexpressive{} is proven correct in the Lean theorem prover and implemented as a practical printer \prettiester{}, which powers a real-world code formatter for the Racket programming language.
Our results show that \prettiester{} (and \Pexpressive{}) is both pretty and fast.
\section*{Data-Availability Statement}
The latest version of the Racket \prettiester{}\footnote{\url{https://github.com/sorawee/pretty-expressive}} and the Racket code formatter\footnote{\url{https://github.com/sorawee/fmt}} are available on GitHub.
The main artifact, which consists of the above softwares and:
\begin{itemize}
\item the Lean formalization (\Cref{sec:semantics,sec:printer})
\item the Rosette proofs (\Cref{sec:tour})
\item the OCaml \prettiester{} (\Cref{sec:impl})
\item the benchmarks to reproduce our evaluation (\Cref{sec:eval})
\end{itemize}
is available on Docker,\footnote{\url{https://hub.docker.com/repository/docker/soraweep/pretty-expressive-oopsla23-artifact/}} with its source on GitHub.\footnote{\url{https://github.com/sorawee/pretty-expressive-oopsla23-artifact}}
A snapshot of the artifact is available on Zenodo~\cite{artifact-doi}.
\if\appendixmode0%
Lastly, the full paper with Appendices A to C is available on arXiv.\footnote{\url{https://arxiv.org/XXXXXX.pdf}}
\fi%

\begin{acks}
We are thankful to the anonymous reviewers and the anonymous artifact reviewers for their very helpful feedback.
This work is supported by the National Science Foundation under Grant Nos. CF-1651225, CCF-1836724, CNS-1844807, and by a gift from the VMware University Research Fund.
\end{acks}

\appendix

\section{An analysis of printers}\label{sec:appendix-survey}

\begin{figure}[H]
\parbox{\figrasterwd}{
\parbox{.6\figrasterwd}{%
\begin{lstlisting}[language=pp]
group (text "AAA" <> nl) <>
nest 5 (group (text "B" <> nl <>
               text "B" <> nl <> text "B"))
\end{lstlisting}
  \vspace{-3mm}
}
\hskip1em
\parbox{.15\figrasterwd}{%
    \begin{lstlisting}[numbers=left, numbersep=.75em, language=literal]
AAA  (!\tikzmark{gwadler-non-optimal-opt-layout1l}!) (&|&)
B B B(!\tikzmark{gwadler-non-optimal-opt-layout2l}!) (&|&)
\end{lstlisting}
\begin{tikzpicture}[remember picture,overlay]
\draw[red, ultra thick, densely dashed]
([shift={(0pt,7pt)}]pic cs:gwadler-non-optimal-opt-layout1l)
--
([shift={(0pt,-2pt)}]pic cs:gwadler-non-optimal-opt-layout2l);
\end{tikzpicture}
    \vspace{-5mm}
}
\hskip1em
\parbox{.15\figrasterwd}{%
    \begin{lstlisting}[numbers=left, numbersep=.75em, language=literal]
AAA B(!\tikzmark{gwadler-non-optimal-layout1l}!) (&|&)
     B(&|&)
     (!\tikzmark{gwadler-non-optimal-layout2l}!)B(&|&)
\end{lstlisting}
\begin{tikzpicture}[remember picture,overlay]
\draw[red, ultra thick, densely dashed]
([shift={(0pt,7pt)}]pic cs:gwadler-non-optimal-layout1l)
--
([shift={(0pt,-2pt)}]pic cs:gwadler-non-optimal-layout2l);
\end{tikzpicture}
    \vspace{-5mm}
}
}
\scriptsize
\caption{A document in the traditional PPL and two of its corresponding layouts. 
Under the width limit of 5, the first layout is optimal---it does not overflow and occupies a minimal number of lines. 
By contrast, the second layout, which is produced by Wadler's printer, overflows and does not occupy a minimal number of lines.
}
\label{fig:wadler-non-optimality}
\end{figure}

\begin{figure}[H]
\centering
\begin{lstlisting}[language=pp]
let rec quadratic (n : int): doc = 
  if n = 0 then text "line"
  else group (quadratic (n - 1) <> nl <> text "line")
\end{lstlisting}
\vspace{-2mm}
\caption{
    The function \pp{quadratic} generates a document of size $O(n)$ that Wadler's algorithm takes $O(n^2)$ to print at any width limit, due to repeated flattening.
}
\label{fig:wadler-time-complexity}
\end{figure}

\begin{figure}[H]
\parbox{.6\figrasterwd}{%
\begin{lstlisting}[language=pp]
text "xxxxxx" <$>
((text "aaa" <+> text "bbb") <|>
 (text "aaa" <$> text "bbb"))
\end{lstlisting}
  \vspace{-3mm}
}
\hskip1em
\parbox{.15\figrasterwd}{%
    \begin{lstlisting}[numbers=left, numbersep=.75em, language=literal]
xxxxx(!\tikzmark{gbernardy-pretty-layout1l}!)x(&|&)
aaa(&|&)
bbb  (!\tikzmark{gbernardy-pretty-layout2l}!) (&|&)
\end{lstlisting}
\begin{tikzpicture}[remember picture,overlay]
\draw[red, ultra thick, densely dashed]
([shift={(0pt,7pt)}]pic cs:gbernardy-pretty-layout1l)
--
([shift={(0pt,-2pt)}]pic cs:gbernardy-pretty-layout2l);
\end{tikzpicture}
    \vspace{-5mm}
}
\hskip1em
\parbox{.15\figrasterwd}{%
    \begin{lstlisting}[numbers=left, numbersep=.75em, language=literal]
xxxxx(!\tikzmark{gbernardy-non-pretty-layout1l}!)x(&|&)
aaabb(!\tikzmark{gbernardy-non-pretty-layout2l}!)b(&|&)
\end{lstlisting}
\begin{tikzpicture}[remember picture,overlay]
\draw[red, ultra thick, densely dashed]
([shift={(0pt,7pt)}]pic cs:gbernardy-non-pretty-layout1l)
--
([shift={(0pt,-2pt)}]pic cs:gbernardy-non-pretty-layout2l);
\end{tikzpicture}
    \vspace{-5mm}
}
\caption{
    A document in the arbitrary-choice PPL and two of its corresponding layouts. 
    Under the width limit of 5, the first layout minimally overflows. 
    By contrast, the second layout, which is produced by Bernardy's practical implementation, overflows more than necessary.
}
\label{fig:bernardy-prettiness}
\end{figure}

\begin{figure}[H]
\centering
\begin{subfigure}[t]{\textwidth}
\begin{lstlisting}[language=pp, mathescape]
let rec mk (n : int): doc = 
  if n = 0 then text "X" <|> text "XX"
  else let subdoc = mk (n - 1) in (chr n <+> subdoc <+> chr n) <|> subdoc
\end{lstlisting}
  \vspace{-2mm}
  \caption{The function \pp{mk} generates a document whose DAG size is $O(n)$. 
  $\mathsf{chr}(n)$ denotes a \pp{text} whose content is a string of length one that contains the $n$th character.}
\end{subfigure}
\begin{subfigure}[t]{\textwidth}
  \input{fig/yelland-sharing-trace.tex}
  \caption{Let $D_n$ denote $\mathsf{mk}(n)$. 
  Yelland's $C'$ function would transform the original document $D_n$ into a restricted document where every aligned concatenation has a \pp{text} as its left subdocument.
  However, the above derivation shows that the transformation has a combinatorial explosion.
  Define $Z_{[]}$ to be $\blacksquare$ in Yelland's paper and $Z_{[x, x_1, \ldots, x_n]}$ to be $C'[\mathsf{chr}(x), Z_{[x_1, \ldots, x_n]}]$.
  The derivation shows that $D_{n-k}$ is recursively transformed in $2^k$ different contexts.
}
\end{subfigure}
\caption{
  A family of documents that illustrates how the transformation $C'$ in Yelland's algorithm does not necessarily preserve the sharing structure in the original document.
}
\label{fig:yelland-sharing}
\end{figure}

\newpage

\begin{figure}[H]
  \centering
  \begin{subfigure}[t]{.58\textwidth}
    \vspace{-2.2in}
\begin{lstlisting}[language=pp]
(* make an empty document of size n; n >= 1 *)
let rec make_dummy (n : int): doc =
  if n = 1 then text ""
  else text "" <+> make_dummy (n - 1)

(* make n lines; n >= 1 *)
let rec make_lines (n : int): doc =
  if n = 1 then text ""
  else text "" <$> make_lines (n - 1)

(* nth triangle number *)
let tri (n : int): int = n * (n + 1) / 2

let make_choices (k : int): doc =
  let rec loop (i : int): doc =
    let doc =
      (make_lines i) <+>
        text (String.make (tri (k - i + 1)) 'a')
    in if i = 1 then doc else doc <|> loop (i - 1)
  in loop k

let rec example (k : int): doc =
  let dummy = make_dummy (k * k) in
  let giant = make_choices k in
  dummy <+> giant
\end{lstlisting}
    \caption{The function \pp{example} produces a document that triggers the worst-case time complexity of Yelland's algorithm (that we are aware of).
    For a fixed $k$, \pp{giant} is a document with $k$ choices, where the $i$-th choice has $i$ lines and $\mathsf{tri}(k - i + 1)$ characters ($\mathsf{tri}$ is the triangle number function).
    Thus, its document tree size is $O(k^2)$.
    By concatenating \pp{giant} with \pp{dummy}, which is an ``empty'' document of size $O(k^2)$, the total document tree size is still $O(k^2)$.
    \pp{giant} is designed so that it has $k$ segmented linear cost functions.
    Thus, the aligned concatenation of \pp{dummy} and \pp{giant} takes $O(k^3)$.
    By normalizing the document size to $\hat{n}$, we obtain that the time complexity of the printer is $O(\hat{n}^{3/2})$.
    }\label{fig:yelland-program}
  \end{subfigure}\hspace{1em}
  \begin{subfigure}[t]{.38\textwidth}
\newcommand*{\ShowIntersection}[3]{
\fill 
    [name intersections={of=#1 and #2, name=i, total=\t}] 
    [red, opacity=1, every node/.style={above left, black, opacity=1}] 
    (i-1) circle (2pt)
        node [below right] {#3};
}

\begin{tikzpicture}
\begin{axis}[xlabel=column position,ylabel=cost,
xmin=-1,xmax=12,ymin=-5,ymax=70, axis lines=center,
width=1.2\textwidth,
height=1.3\textwidth,
cycle list={[colors of colormap={0,100,...,1000}]}]
\addplot+[domain=-10:20,name path global=p1]{10*x+1};
\addplot+[domain=-10:20,name path global=p2]{9*x+3};
\addplot+[domain=-10:20,name path global=p3]{8*x+6};
\addplot+[domain=-10:20,name path global=p4]{7*x+10};
\addplot+[domain=-10:20,name path global=p5]{6*x+15};
\addplot+[domain=-10:20,name path global=p6]{5*x+21};
\addplot+[domain=-10:20,name path global=p7]{4*x+28};
\addplot+[domain=-10:20,name path global=p8]{3*x+36};
\addplot+[domain=-10:20,name path global=p9]{2*x+45};
\addplot+[domain=-10:20,name path global=p10]{1*x+55};
\ShowIntersection{p1}{p2}{2}
\ShowIntersection{p2}{p3}{3}
\ShowIntersection{p3}{p4}{4}
\ShowIntersection{p4}{p5}{5}
\ShowIntersection{p5}{p6}{6}
\ShowIntersection{p6}{p7}{7}
\ShowIntersection{p7}{p8}{8}
\ShowIntersection{p8}{p9}{9}
\ShowIntersection{p9}{p10}{10}
\end{axis}
\end{tikzpicture}

    \caption{A plot of the piecewise linear cost function (lines along the red dots) for \pp{giant} in \Cref{fig:yelland-program} with $k = 10$.
    The x-axis is column positions at which \pp{giant} will be printed.
    The y-axis is cost of \pp{giant}.
    The plot consists of $O(k)$ segmented cost functions, where each segment is a linear function.
    For simplicity, we assume that (1) the page width limit is $0$; (2) there is no cost for newlines; and (3) the cost for each character past the page width limit is $1$.
    Let $\cl{d}_i$ be the $i$-th choice in \pp{giant}.
    The cost function for $\cl{d}_i$ then is $C_{\cl{d}_i}(c) = ic + \mathsf{tri}(k - i + 1)$.
    These cost functions intersect at $c = 2, \ldots, k$.
    Thus, the cost function for \pp{giant} is unable to prune any segments away.}\label{fig:yelland-time-complexity-plot}
  \end{subfigure}
  \caption{
      In Yelland's algorithm, every choiceless document (in the arbitrary-choice PPL) $\cl{d}$ has an associated \emph{piecewise linear} cost function $C_{\cl{d}}$, where
      $C_{\cl{d}}(c)$ determines the cost of $\cl{d}$'s rendered layout at the column position $c$.
      A general document $d$ similarly has an associated piecewise linear cost function $C_{d}$, which takes the minimum of the cost functions from all choiceless documents that $d$ generates.
      The algorithm appears to be efficient at first glance, since taking the minimum can prune away many segmented linear cost functions.
      However, we are able to construct a document \pp{giant} of size $O(\hat{n})$ whose cost function has $O(\sqrt{\hat{n}})$ segmented linear cost functions, where $\hat{n}$ is the tree size of the document.
      As the time complexity of the printer is $O(\hat{n} M)$ where $M$ is the maximum number of piecewise linear cost functions in a cost function, 
      we obtain $O(\hat{n}^{3/2})$.
  }
  \label{fig:yelland-time-complexity}
\end{figure}

\section{Selected proof sketches}\label{sec:proofs}

\arbitrarylexpressivefunctional*

\begin{proofsketch}
    For the arbitrary-choice PPL with the evaluation function $\render(\cdot)$, let $L$ be any non-empty set of layouts. 
    For each $l_i \in L$ where $l_i = [s^i_1, \ldots, s^i_{|l_i|}]$, we construct $d_i$ to be $\text{\pp{text} } s^i_1 \text{ \pp{<$>} } \ldots \text{ \pp{<$> text} } s^i_{|l_i|}$. 
    Finally, we construct $d$ to be \lstinline[language=pp, mathescape]{$d_1$ <|> $\ldots$ <|> $d_{|L|}$}.
    We can see that $\render(d) = L$.
    The proof for \Lexpressive{} PPL is similar, but we replace $a \text{ \pp{<$>} } b$ with $a \text{ \pp{<> nl <>} } b$.
\end{proofsketch}

\newpage

\traditionalfunctional*

\begin{proofsketch}
    It is not possible to construct a document in the traditional PPL that evaluates to the set of layouts 
    $E = \set{[\text{\pp{"a"}}], [\text{\pp{"b"}}]}$.
    To see why, let $\mathsf{rmspace} : \LayoutT \to \StrT$ be a function that joins all lines in a layout into a single line, with all whitespaces removed, and lift $\mathsf{rmspace}$ to work on a set of layouts (i.e., $\mathsf{rmspace}(L) = \set{\mathsf{rmspace}(l) : l \in L}$.
    Let $\render(\cdot)$ be the evaluation function for the traditional PPL.
    We can prove by induction that $\mathsf{rmspace}(\render(d))$ is a singleton set for any document $d$. 
    In other words, all layouts in $\render(d)$ are the same, modulo whitespaces.
    However, $\mathsf{rmspace}(E) = \set{\text{\pp{"a"}}, \text{\pp{"b"}}}$, which is not a singleton set. 
    Hence, by congruence, no document can render to $E$.

    Note that there are other sets of layouts that are the same modulo whitespaces, but can't be evaluated to by the traditional PPL.
    An example is \emph{synchronized} differences of spacing across multiple lines.
\end{proofsketch}

\sublexpressivenotfunctional*

\begin{proofsketch}
    It is not possible to construct a document in each language in question that evaluates to the following set of layouts
    \paragraph{\Lexpressive{} without \pp{text}}
    $\set{[\text{\pp{"a"}}]}$, because all we can produce is whitespaces.
    \paragraph{\Lexpressive{} without \pp{<>}}
    $\set{[\text{\pp{"a"}}, \text{\pp{"b"}}, \text{\pp{"c"}}]}$, because all we can produce is at most two lines.
    \paragraph{\Lexpressive{} without \pp{nl}}
    $\set{[\text{\pp{"a"}}, \text{\pp{"b"}}]}$, because all we can produce is a single line.
    \paragraph{\Lexpressive{} without \pp{<|>}}
    $\set{[\text{\pp{"a"}}], [\text{\pp{"b"}}]}$, because all we can produce is a single layout.
\end{proofsketch}

\lexpressiveexpressibility*

\begin{proofsketch}
    The following syntactic abstractions can be used to define the constructs:

    \begin{itemize}
        \item \pp{group} is definable by $\mathbf{M}(\alpha_1) = \alpha_1\ \text{\pp{<|> flatten}}\ \alpha_1$ 
        \item \pp{<$>} is definable by $\mathbf{M}(\alpha_1, \alpha_2) = \alpha_1\ \text{\pp{<> nl <>}}\ \alpha_2$.
        \item \pp{<+>} is definable by $\mathbf{M}(\alpha_1, \alpha_2) = \alpha_1\ \text{\pp{<> align}}\ \alpha_2$.
    \end{itemize}

    The rest of the constructs are already in \Lexpressive{}.
\end{proofsketch}

\inexpressibility*

\begin{proofsketch}
    Let $\render_a(\cdot)$ and $\render_b(\cdot)$ denote the evaluation functions for $\Sigma$ and $\Sigma \cup \set{\mathbf{F}}$, respectively.
    We prove the contraposition. 
    Assuming that $\mathbf{F}$ is definable in $\Sigma$, we need to prove that for any $d_1$, $d_2$, and $R$, \obsequiv{R}{\Sigma}{d_1}{d_2} implies \obsequiv{R}{\Sigma \cup \set{\mathbf{F}}}{d_1}{d_2}.
    Let $d_1$, $d_2$, and $R$ be arbitrary.
    We suppose that for all context $C$ in $\Sigma$, $R(\render_a(C(d_1)), \render_a(C(d_2)))$ holds, and need to prove that for all context $C$ in $\Sigma \cup \set{\mathbf{F}}$, $R(\render_b(C(d_1)), \render_b(C(d_2)))$ holds.

    Let $C$ be a context in $\Sigma \cup \set{\mathbf{F}}$.
    Because $\mathbf{F}$ is definable in $\Sigma$, we can perform a syntactic expansion on $C$ to obtain a context $C^*$ in $\Sigma$ such that $\render_a(C^*(d)) = \render_b(C(d))$ for all document $d$ in $\Sigma$.
    Hence, it suffices to prove that $R(\render_a(C^*(d_1)), \render_a(C^*(d_2)))$ holds, but this is our hypothesis (instantiated with $C^*$).
\end{proofsketch}

\inexpressibilityoldlangs*

\begin{proofsketch}
    In each proof, we need to show that $\mathbf{F}$ is not definable in $\Sigma$, where $\mathbf{F}$ and $\Sigma$ are the construct and the PPL in question.
    We do so by providing a counterexample, which consists of documents $d_1$ and $d_2$, and the relation $R$.
    By induction, it can be shown that \obsequiv{R}{\Sigma}{d_1}{d_2}.
    We will further provide a counterexample context to show that \notobsequiv{R}{\Sigma \cup \set{\mathbf{F}}}{d_1}{d_2}.
    By \Cref{thm:inexpressibility}, this suffices to show that $\mathbf{F}$ is not definable in $\Sigma$.

    \paragraph{\pp{<>} is not definable in the arbitrary-choice PPL}
    Given $\mathsf{maxWidth}$ from \Cref{ex:gen-equiv}, the counterexample is $d_1 = \text{\pp{text "a" <$> text "bb"}}$, $d_2 = \text{\pp{text "aa" <$> text "bb"}}$, and $R = \set{(L_a, L_b): \mathsf{maxWidth}(L_a) = \mathsf{maxWidth}(L_b)}$.
    In particular, with $C(\alpha) = \text{\pp{text "c" <>}}\ \alpha$, 
    we have that \\
    $\mathsf{maxWidth}(\render(d_1)) = \set{2}$, but $\mathsf{maxWidth}(\render(d_2)) = \set{3}$.

    \paragraph{\pp{nest} is not definable in the arbitrary-choice PPL}
    Given $\mathsf{maxWidth}$ from \Cref{ex:gen-equiv}, the counterexample is
    $d_1 = \text{\pp{text "bb" <$> text "a"}}$, $d_2 = \text{\pp{text "cc" <$> text "bb" <$> text "a"}}$, and $R = \set{(L_a, L_b): \mathsf{maxWidth}(L_a) = \mathsf{maxWidth}(L_b)}$.
    In particular, with $C(\alpha) = \text{\pp{nest}}\ 1\ \alpha$, 
    we have that $\mathsf{maxWidth}(\render(d_1)) = \set{2}$, but $\mathsf{maxWidth}(\render(d_2)) = \set{3}$.

    \paragraph{\pp{group} is not definable in the arbitrary-choice PPL}
    Let $\mathsf{maxa} : \LayoutT \to \mathbb{N}$ be a function that finds the maximum number of the character ``a'' in lines of the layout, and lift $\mathsf{maxa}$ to work on a set of layouts.
    The counterexample is $d_1 = \text{\pp{text "a" <$> text "a"}}$, $d_2 = \text{\pp{text "a" <$> text "a" <$> text "a"}}$, and $R = \set{(L_a, L_b): \mathsf{maxa}(L_a) = \mathsf{maxa}(L_b)}$.
    In particular, with $C(\alpha) = \text{\pp{group}}\ \alpha$, 
    we have that $\mathsf{maxa}(\render(d_1)) = \set{1, 2}$, but $\mathsf{maxa}(\render(d_2)) = \set{1, 3}$.

    \paragraph{\pp{<+>} is not definable in the traditional PPL}
    Let $\mathsf{spaces} : \LayoutT \to \mathbb{N}$ be a function that counts the number of spaces in a layout (not counting newlines), and lift $\mathsf{spaces}$ to work on a set of layouts.
    The counterexample is $d_1 = \text{\pp{text "a"}}$, $d_2 = \text{\pp{text "aa"}}$, and $R = \set{(L_a, L_b): \mathsf{spaces}(L_a) = \mathsf{spaces}(L_b)}$.
    In particular, with $C(\alpha) = \alpha\ \ \text{\pp{<+> (text "b" <> nl <> text "c")}}$, 
    we have that $\mathsf{spaces}(\render(d_1)) = \set{1}$, but $\mathsf{spaces}(\render(d_2)) = \set{2}$.
\end{proofsketch}

\functionalityimpliesinexpressibility*

\begin{proofsketch}
    Because $\Sigma$ is not functionally complete, there is a set of layouts $L^*$ that can't be evaluated to by any document in $\Sigma$.
    Since $\Sigma \cup \set{\mathbf{C}}$ is functionally complete, there is a document $d^*$ (which necessarily contains $\mathbf{C}$) that evaluates to $L^*$.
    Let $d_1$ and $d_2$ be any document in $\Sigma$, and $R = (2^{\LayoutT} \times 2^{\LayoutT}) \setminus \set{(L^*, L^*)}$.
    Then \obsequiv{R}{\Sigma}{d_1}{d_2} holds trivially.
    However, with $C(\alpha) = d^*$,
    we have that \notobsequiv{R}{\Sigma \cup \set{\mathbf{C}}}{d_1}{d_2}. 
    This concludes the proof that $\mathbf{C}$ is not definable in $\Sigma$.
\end{proofsketch}

\minimal*

\begin{proofsketch}
    The proofs for \pp{text}, \pp{nl}, \pp{<>}, and \pp{<|>} are applications of \Cref{lem:arbitrary-lexpressive-functional}, \Cref{lem:sub-lexpressive-not-functional}, and \Cref{lem:functionality-implies-inexpressibility}.
    The proofs for \pp{nest}, \pp{flatten}, and \pp{align} are just like how we proved \Cref{thm:inexpressibility-old-langs} for \pp{nest}, \pp{group}, and \pp{<+>}.
\end{proofsketch}

\timecomplexity*

\begin{proofsketch}
  The most expensive operation in the printer is concatenation (via \textsc{ConcatRSSet}).
  The operation resolves the left sub-document, resulting in a measure set whose size is at most $\cfw$ according to \Cref{lem:number-of-measures}.
  It then resolves the right sub-document in at most $\cfw$ different contexts.
  Thus, there are at most $\cfw^2$ different measures from the right sub-document that the printer needs to concatenate and prune. 

  Consider $\pL{d}{c}{i}{S}$.
  $d$ can range over $n$ different values.
  $c$ and $i$ can range over $\cfw$ different values that are under $\cfw$.
  Hence, there are $O(n \cfw^2)$ different contexts under the computation width limit.
  Multiplying this with the maximum units of computation in the previous paragraph, 
  we obtain that the time complexity due to resolving within $\cfw$ is $O(n \cfw^4)$, assuming that the resolver reuses memoized measure sets under the same context.

  When $d$ is printed beyond $\cfw$, however, it can be fully resolved at most once, because:
  \begin{enumerate}
    \item While we would resolve both sub-documents of choice nodes, they would be all tainted, due to \Cref{lem:print-beyond-limit}. 
    Because all tainted measure sets are promises, all computations are delayed.
    The merge operation then chooses only one tainted measure set as the result, discarding the other one.
    \item The document is properly shared, so under a given path, a document is encountered at most once.
  \end{enumerate}
  As a result, the time complexity due to printing over $\cfw$ is simply $O(n)$.
  Combining both parts, we obtain that the time complexity of \Pexpressive{} is $O(n \cfw^4)$.
\end{proofsketch}

\timecomplexityarbfragment*

\begin{proofsketch}
  In the arbitrary-choice PPL, $c = i$ is (mostly) maintained throughout the printing.
  Hence, there is one less dimension to consider, leading to the time complexity of $O(n \cfw^3)$.
\end{proofsketch}

\section{Discussion}\label{sec:discussion}

In this section, we broadly discuss the design of our work.

\subsection{Additional Constructs}

\prettiester{} supports additional constructs \pp{fail}, \pp{newline}, and \pp{reset}.
The Racket \prettiester{} further supports additional constructs \pp{full} and \pp{cost}.
These constructs are out of scope for the paper, and we leave their formalization as future work.

\paragraph{Failure}
\pp{fail} widens to the empty set, thus introducing the possibility that a printing could fail. 
Furthermore, it is the identity for the operation \pp{<|>}.
\pp{fail} makes \Lexpressive{} more expressive because it is impossible to make a document in \Lexpressive{} evaluate to the empty set.
In this sense, it could be said that \Lexpressive{} is not truly ``functionally complete,'' but \Lexpressive{} with \pp{fail} is.
Supporting \pp{fail} can be done via rewriting rules:
every document with \pp{fail} can be normalized to a semantically-equivalent document without \pp{fail}, or to a single \pp{fail}.
Hence, there is no need to modify the core printer to support the construct.

\paragraph{Generalized Newlines}
$\text{\pp{newline} } m$ is a straightforward generalization of \pp{nl} so that flattening it can result in other possibilities besides a single space.
When $m$ is $\text{\pp{Some} } s$, the flattened result is $\text{\pp{text} } s$.
When $m$ is \pp{None}, the flattened result is \pp{fail}.
With this construct: 

\begin{itemize}
\item \pp{nl} is definable with \pp{newline (Some " ")}. 
\item \pp{break} from \citet{leijen}'s printer is definable with \pp{newline (Some "")}.
\item \pp{hard_nl} is definable with \pp{newline None}. 
With \pp{hard_nl}, \pp{singleLine} from \citet{bernardy:implementation}'s practical printer is definable with \pp{flatten}, given that the vertical concatenation uses \pp{hard_nl} for entering a newline.
\end{itemize}

\paragraph{Reset}

$\text{\pp{reset} } d$ resets the indentation level to 0 for $d$.
This is useful for formatting multiline comments and here-string.

\paragraph{Fullness}

$\text{\pp{full} } d$ marks $d$ as \emph{full}, which means there must be no more text after it in the same line.
The construct is especially useful for formatting line comments, as it is illegal to put a piece of code after a line comment.
A simpler variant of \pp{full} is also implemented in Yelland's printer for the R code formatter~\cite{r:fmt}.
Unlike other extensions, which can be supported without significant changes to the core printer modification, \pp{full} requires more involved changes.
\begin{itemize}
    \item 
    The measure set definition is now required to recognize the empty set (where we prefer a tainted measure set over the empty set).
    \item 
    The resolver would consume two additional boolean arguments, which indicate the fullness status before and after the document.
    \item 
    Merging two tainted measure sets must now keep both tainted measure sets, 
    and we may need to try both if the first one resolves to the empty set.
\end{itemize}

To keep the time complexity of the printer $O(n\cfw^4)$, we rely on the fact that ``emptiness'' in resolving (that is, resolving to the empty set) is independent from column positions and indentation levels. 
Thus, even though we now need to try many tainted measure sets, a document can be tried at most four times, which bounds the time complexity.

\paragraph{Cost}
$\text{\pp{cost} } \mathcal{C}\ d$ adds a cost $\mathcal{C}$ to measures due to $d$.
This construct is not expressive in the traditional sense, as it does not affect layout results. 
However, it allows us to make \emph{weighted} choices, so that we can prefer one style over another when all else is equal.
Due to the flexibility of the cost factory, it is even possible to make multidimensional weights. 

\subsection{Safety}
As shown in the proof of \Cref{lem:traditional-functional}, the traditional PPL is not functionally complete because all layouts must have the same content, modulo whitespaces.
While this property is restrictive for many tasks as elaborated in the paper, it does provide a sort of safety guarantee that the layouts will not be wildly different.
\pp{<|>}, however, allows us to violate this property.
In fact, some arbitrary-choice printers (e.g. a prototype of Bernardy's printer~\cite{bernardy:prototype}) \emph{intend} that \pp{<|>} should be restricted to maintain the property.
Similarly, the inclusion of \pp{fail}, \pp{newline}, or \pp{full} makes it possible to evaluate to an empty set, but the PPLs without the essence of \pp{fail} provide a safety guarantee that an evaluation will never result in an empty set.
Generally, the more expressive a language is, the more properties it will break, and the more burden will be put on the users to carefully use the constructs.

We argue that the spirit of these safety properties can still be accomplished in PPLs with a functionally complete core.
One possible approach is similar to Wadler's treatment of \pp{<|>} and \pp{group}: define high-level, ``safe'' constructs with just enough expressiveness to solve a domain-specific task, based on the core, ``unsafe'' constructs, and then hide these ``unsafe'' constructs away from the external interface.
For example, one may hide \pp{<|>}, and instead provide $\mathsf{groupParen}(d) = \text{\pp{(text "(" <>} } d \text{ \pp{<> text ")") <|> flatten} } d$, which evaluates to either $d$ with parentheses wrapped around, or the flattened $d$.
The language as defined by the external interface is no longer functionally complete, but enjoys the property that all layouts are the same modulo whitespaces and parentheses. 
Another possible approach is to export the core, ``unsafe'' constructs, but perform a static analysis to ensure that the document satisfies intended safety properties.

In any case, the expressive core constructs are what enable the advanced features that languages may require to be rendered well.
Thus, our view is that an expressive printer is the key. 
We should start with an expressive albeit unsafe printer, rather than a safe but non-expressive one.

\subsection{Memoization}
While memoization is important to guarantee that \Pexpressive{} will not take exponential time, it is also the performance bottleneck when the input document is large, due to too much memory allocation.
In \prettiester{}, we employ a heuristic to reduce memory allocation
by adding a metadata \emph{memoization weight} to each document node, which counts how long memoization has not been performed on descendant nodes.
When the weight reaches a limit (set to \memoizationLimit{} in our implementation), we perform memoization on the node, and reset the weight to 0.
This can significantly speed up the performance of \prettiester{} on some large documents.

\subsection{Fusing Resolving and Rendering}
One optimization in \prettiester{} is to fuse together the resolving of a document to a measure set and the rendering of a choiceless document to a layout. 
This is done by replacing the $\mL$ component in a measure with a \emph{token function}, which consumes a list of rendered tokens \emph{after} the document is placed, and returns a new list of rendered tokens. 
A similar technique was employed by \citet{podkopaev}.

\subsection{Handling Bias in the Presence of Taintedness}

In \Cref{sec:printer}, we see that the merge operation and thus the \pp{<|>} operator is left-biased in the presence of taintedness.
When exceeding \cfw{} is unavoidable,
all text could be put in one line in the worst case if all left sub-documents use the ``horizontal styling''!
The proper solution is to increase \cfw{}.
However, \prettiester{} also implements a heuristic to infer a sub-document with the ``vertical styling.''
The heuristic adds a metadata that overestimates the number of lines for each document node. 
\prettiester{} then uses a document with a larger overestimated number of lines as the left sub-document in choice documents.

\subsection{Partial Evaluation}

Similar to how we can perform partial evaluation in programming languages, we can also perform partial evaluation in PPL using rewriting rules.
For example, a concatenation of two \pp{text} can immediately be partially evaluated to a single \pp{text}.
However, this partial evaluation must be done with care to still preserve the sharing structure, since unconstrained rewriting may unfold the DAG structure into a tree, as illustrated in \Cref{fig:yelland-sharing}. 
It is also worth noting that the partial evaluation may not necessarily preserve the semantics in the presence of taintedness. 
For example, one may want to reduce a \lstinline[language=pp, mathescape]{nest $n$ (text $s$)} to \lstinline[language=pp, mathescape]{text $s$} for any $n$ and $s$, but when $n > \cfw$, the document will definitely resolve to a tainted measure set, while the partially evaluated one does not necessarily.\footnote{One may argue, however, that this semantic change is acceptable, because the change is for the better.}

\bibliography{paper}

\end{document}